# Nanoelectromechanics of Piezoresponse Force Microscopy

Sergei V. Kalinin

Condensed Matter Sciences Division, Oak Ridge National Laboratory,

Oak Ridge, TN 37831

e-mail: sergei2@ornl.gov

Edgar Karapetian

Department of Mathematics & Computer Science

Suffolk University, Boston, MA 02114

e-mail: karap@comcast.net

Mark Kachanov

Department of Mechanical Engineering

Tufts University, Medford, MA 02155

e-mail: Mark.Kachanov@tufts.edu

## Abstract

To achieve quantitative interpretation of Piezoresponse Force Microscopy (PFM), including resolution limits, tip bias- and strain-induced phenomena and spectroscopy, analytical representations for tip-induced electroelastic fields inside the material are derived for the cases of weak and strong indentation. In the weak indentation case, electrostatic field distribution is calculated using image charge model. In the strong indentation case, the solution of the coupled electroelastic problem for piezoelectric indentation is used to obtain the electric field and strain distribution in the ferroelectric material. This establishes a complete continuum mechanics description of the PFM contact mechanics and imaging mechanism. The electroelastic field distribution allows signal generation volume in PFM to be determined. These rigorous solutions are compared with the electrostatic point charge and sphere-plane models, and the applicability limits for asymptotic point charge and point force models are established. The implications of these results for ferroelectric polarization switching processes are analyzed.

## I. Introduction

Progress in oxide electronic devices including microelectromechanical systems (MEMS), nonvolatile ferroelectric memories (FeRAMs), and ferroelectric heterostructures requires an understanding of local ferroelectric properties at the nanometer level. This has motivated an increasing number of studies of ferroelectric materials with various scanning probe microscopies (SPM).[1,2,3,4] Among the techniques for local ferroelectric imaging, the most widely used currently is Piezoresponse Force Microscopy (PFM), due to the ease of implementation, high resolution, and its relative insensitivity to topography. PFM is rapidly becoming one of the primary characterization tools in the ferroelectric thin film research that routinely allows high resolution (~3 - 10 nm) domain imaging. Applications of PFM include imaging static domain structures in thin film, single crystals, and polycrystalline materials; selective poling of specified regions on ferroelectric surface, studies of temporal and thermal evolution of domain structures, and quantitative measurements of thermal phenomena and local hysteresis measurements. For most of these applications, qualitative interpretation of the PFM image in terms of ferroelectric domain morphology is sufficient and detailed knowledge of the PFM imaging mechanism is not required.

In the last several years, significant attention was attracted to quantitative studies of local ferroelectric behavior by PFM. The early applications include PFM voltage spectroscopy, i.e., local hysteresis loop measurements.[5,6] PFM spectroscopy allows ferroelectric properties of the individual grains to be addressed, including remanent response and coercive bias, on the ~50 nm level. Application of high voltage to the tip allows local polarization switching, providing an approach to engineer and control domain structures at the nanoscale. This approach can potentially be used for high-density ferroelectric storage;[7,8]

alternatively, polarization dependent reactivity of the surface in the acid etching[9] or metal photodeposition processes[10,11] can be used to engineer nanoscale structures (ferroelectric lithography). Independently, quantitative PFM measurements were used to address the depth dependence of ferroelectric properties in the beveled thin film structures[12] and ferroelectric size effect in nanocrystals.[13,14] Recently, it was shown that mechanical strain produced by the tip can suppress local polarization[15] or induce local ferroelectroelastic polarization switching.[16,17,18] A quantitative analysis of tip-induced potential and stress distribution in the material is required to characterize local ferroelectric properties by SPM including hysteresis measurements, stress effects in thin films,[19] size dependence of ferroelectric properties,[20,21] bias- and stress-induced polarization switching.

In order to achieve the quantitative understanding of PFM nanoelectromechanics, we analyze tip-induced field distributions for the case of $c^+$, $c^-$ domains in tetragonal perovskite ferroelectrics. For small indentation forces, the electroelastic contribution to the electric field below the tip can be neglected, since the contact area between the tip and the surface is small. In this weak indentation limit, the electric field in the material is calculated using the electrostatic sphere-plane model. It is shown that, under typical PFM imaging conditions, the capacitance of the contact area can be comparable or larger than that of the spherical part of the tip; hence field analysis even in the purely electrostatic case requires the contact area contribution to be taken into account. This is further corroborated by high (~3-10 nm) spatial resolution achievable in PFM which is significantly better than typical radius of curvature for metal coated tip (~ 50 nm) and is comparable to expected tip-surface contact area (~ several nanometers). The rigorous analysis of the field distributions requires solution of the coupled electroelastic problem of the spherical indentation of a piezoelectric that simultaneously takes into account electrostatic, elastic, and electroelastic phenomena at the tip-surface junction. The use of Fabrikant's results in the potential theory,[22,23] coupled with the recently established correspondence principle[24] yields exact solutions in terms of elementary functions for the full field components inside the material. These solutions are compared to the simplified models, and the applicability limits for asymptotic point-charge behavior are established. The implications of these solutions for ferroelectric polarization switching processes are analyzed.

## II. Principles of PFM

PFM is based on the detection of bias-induced surface deformation. The tip is brought into contact with the surface, and the piezoelectric response of the surface is detected as the first harmonic component, $A_{1\omega}$, of the bias-induced tip deflection, $d = d_0 + A_{1\omega}\cos(\omega t + \varphi)$, when the periodic bias, $V_{\mathrm{tip}} = V_{\mathrm{dc}} + V_{ac}\cos(\omega t)$, is applied to the tip. The phase of the electroelastic response of the surface, $\varphi$, yields information on the polarization direction below the tip. For $c^-$ domains (polarization vector pointing downward) the application of a positive tip bias results in the expansion of the sample and surface oscillations are in phase with the tip voltage, $\varphi = 0$. For $c^+$ domains, $\varphi = 180°$. The piezoresponse amplitude, $A = A_{1\omega}/V_{ac}$, defines the local electromechanical activity of the surface. One of the major complications in PFM is that both long-range electrostatic forces and the electroelastic response of the surface contribute to the PFM signal so that the experimentally measured piezoresponse amplitude is $A = A_{el} + A_{piezo} + A_{nl}$, where $A_{el}$ is the electrostatic contribution, $A_{piezo}$ is the electroelastic contribution and $A_{nl}$ is the non-local contribution due to capacitive cantilever-surface interactions.[25,26] Quantitative PFM imaging requires $A_{piezo}$ to be maximized to achieve a predominantly electroelastic contrast. The cantilever size is usually significantly larger than the domain size; therefore, a non-local cantilever contribution is usually present in the form of an additive offset to the PFM image.

Even under optimal conditions, the origins of the electroelastic contribution, $A_{piezo}$, and its relationship to materials properties are not straightforward because of the complex geometry of the tip-surface junction. Some progress in the quantitative understanding of PFM was achieved recently.[27,28,29,30] Depending on the tip radius of curvature and the indentation force, the PFM signal may correspond to the electroelastic response of the surface induced by the contact area (strong indentation limit) or be dominated by the electroelastic response of the surface due to the field produced by the spherical part of the tip (weak indentation limit) as illustrated in Fig. 1 a,b. In these cases, the magnitude of surface and tip displacements is determined by the electromechanical coupling in the material. Alternatively, the signal can be dominated by the electrostatic tip-surface interactions (electrostatic limit) that will result in indentation even for non-piezoelectric materials. Quantitative measurement of the electroelastic properties of the surface is possible only in the strong indentation limit corresponding to a large tip-surface contact area. The measured piezoelectric response in this case is directly related to the piezoelectric constant tensor $d_{ij}$ of the material. However, despite the progress in the interpretation of PFM, little is known about the potential and stress distribution inside ferroelectric materials during imaging. Understanding of these parameters is vital for the interpretation of the PFM hysteresis loops, predicting the PFM resolution limit, and the minimal size of domains and structures that can be patterned by PFM lithography, establishing the relative importance of bias- vs. stress-induced effects, and estimating the degree of invasiveness of technique.

## III. Simplified Models

The potential distribution inside a ferroelectric thin film or crystal was analyzed by a number of authors using a rigid dielectric model that ignores electroelastic coupling in the ferroelectric. Several groups have used point charge or two point-charge models, in which the tip is represented by a point charge $q$ located at a distance $h$ from the surface. The magnitude of the charge and charge-surface separation is selected such that the radius of curvature of isopotential lines and the potential at the surface coincide with the corresponding characteristics of the tip. These models can be readily extended to describe the electrostatics of thin films (as opposed to bulk ferroelectrics) using the set of image charges.[28]

An alternative approach for modeling capacitive tip-surface interactions is based on approximation of the realistic tip shape by suitably chosen simple geometrical shape. A number of geometric models have been used to approximate capacitive tip-surface interactions including a sphere[31], a hyperboloid[32,33,34], a cone[35] or a cone with spherical apex[36] that account for tip apex and conical part of the tip. In addition, cantilever contribution to the overall tip-surface capacitance can be approximated using tilted plane-plane capacitor model.[37]

However, it is recognized that fields produced by both conical part of the tip and the cantilever are non-local and vary on the length scales of several microns (cone) and tens of microns (cantilever), which is significantly large than typical resolution of ~ 10 nm observed in PFM or minimal domain radius (~ 20 nm)[8] that can be produced by local switching. Thus, electrostatic tip-surface interactions can be best modeled using geometric models in which a conductive tip is represented by a conducting sphere touching (tip surface separation $d = 0$) or slightly above ($d > 0$) the ferroelectric surface. In-depth analysis of field distribution and domain switching processes using these models was given by Molotskii *et al.*,[38,39] and independently by Abplanalp.[40] To establish the validity of these electrostatic models, we now analyze the applicability of the point-charge model compared to the full electrostatic sphere-plane model and estimate the contribution of the contact area to the capacitance of the tip-surface system and hence to the electrostatic field inside the material.

In the electrostatic sphere-plane model, the potential inside the ferroelectric is approximated using a point-charge model with charge $C_d V$ located at a distance $R$ from the surface, where $V$ is the tip bias and $C_d$ is the conductive sphere-dielectric plane capacitance,[41,42]

$$C_d(\kappa)_{z=0} = 4\pi\varepsilon_0 R \frac{\kappa+1}{\kappa-1} \ln\left(\frac{\kappa+1}{2}\right), \tag{1}$$

where $R$ is the radius of curvature of the tip and $\kappa$ is the dielectric constant. For an anisotropic dielectric material, the effective dielectric constant, $\kappa = \sqrt{\kappa_{11}\kappa_{33}}$ , where $\kappa_{11}$, $\kappa_{33}$ are the principal values of the dielectric constant tensor. This approximation, which neglects the contribution of the contact area to the tip-surface capacitance and hence to the potential inside the material, is appropriate in the weak indentation limit. However, it has been shown that quantitative imaging and polarization switching in the ferroelectric materials occurs predominantly in the strong indentation limit, in which a substantial contact between the tip and the surface is established.[27] Under these conditions, the contribution of the contact area to tip-surface capacitance can become comparable to the sphere-plane capacitance and corresponding equivalent circuit is illustrated in Fig. 1 c. The capacitance of the contact area can be estimated as

$$C_{ca} = 4\kappa\varepsilon_0 a, \tag{2}$$

where $a$ is radius of contact. In both sphere-plane and disc-plane model in Eqs.(1,2) the tip surface has uniform potential, whereas corresponding induced charge density is highly non-uniform. The applicability of Eq. (2) for small contact radii is limited by the quantum capacitance of the junction, in which case the Thomas-Fermi length in metallic tip or Debye length in semiconducting tip can be comparable to the contact radius, resulting in the significant potential drop in the junction region and decreasing overall contact capacitance. Corresponding quantum capacitance can be estimated as $C_q = \kappa_c \varepsilon_0 \pi a^2 / \lambda$, where $\kappa_c$ is the effective dielectric constant of the contact material and $\lambda$ is the thickness of contact region. For ideal contact, $\lambda$ is of the order of magnitude of Thomas-Fermi length for metal (0.5 – 1 Å) or characteristic extrapolation length for ferroelectric. The tip surface contact capacitance Eq. (2) and quantum capacitance are connected in series (Fig. 1 c), suggesting that the quantum correction to capacitance is significant when $C_q = C_{ca}$, corresponding to $a = 1.273\lambda(\kappa/\kappa_c)$. Depending on the ratio between the bulk dielectric constant of ferroelectric and dielectric constant of the contact layer, the quantum capacitance can limit overall contact capacitance for contact radii as large as several nanometers and rigorous analysis of this behavior requires atomistic simulation of electrostatic and dielectric properties of metal-ferroelectric interface. However, in this regime contact capacitance is also expected to be dominated by sphere-plane capacitance, as illustrated below. The total field distribution produced by contact area and spherical and conical parts of the tip and cantilever can be represented as shown in Fig. 1 d. Note that only contact area and spherical parts of the tip provide field distributions localized enough to account for observed PFM resolution, whereas fields from conical part of the tip and cantilever will produce position-independent constant offset to PFM signal.

As follows from Eqs. (1,2), capacitance scales logarithmically with the dielectric constant of the substrate for the spherical part of the tip and linearly for the contact area, while both scale linearly with the corresponding radii. The critical ratio $\eta_{crit} = R/a$ of the tip radius, $R$, to the contact radius, $a$, for which the corresponding capacitances are equal, $C_{ca} = C_d$, can be calculated as a function of dielectric constant (Fig. 2a). From these simple estimates, the effective tip radius for most ferroelectric materials must be at least 1-1.5 orders of magnitude larger than the contact radius for the tip capacitance contribution to dominate. These estimates can be further extended using the Hertzian contact model to relate tip radius and contact diameter. The relationship between the indentation depth, $w_0$, tip radius of curvature, $R$, and load, $P$, is[43]

$$w_0 = \left(\frac{3P}{4E^*}\right)^{\frac{2}{3}} R^{-\frac{1}{3}} \tag{3}$$

where $E^*$ is the effective Young's modulus of the tip-surface system defined as

$$\frac{1}{E^*} = \frac{1-\nu_1^2}{E_1} + \frac{1-\nu_2^2}{E_2} \tag{4}$$

$E_1$, $E_2$ and $\nu_1$, $\nu_2$ are Young's moduli and Poisson ratios of tip and surface materials. For ferroelectric perovskites, Young's modulus $E^*$ is of the order of 100 GPa. The elastic modulus of the tip can vary significantly depending on the material used. For hard conductive coatings such as TiN, $W_2C$ and doped diamond, Young's modulus is of order of 400 – 1000 GPa depending on deposition conditions and therefore tip deformation during the indentation process can be neglected. For doped silicon ($E_{Si}$ = 107 GPa) tips and particularly for tips coated by conductive metals such as Au or Pt ($E_{Au}$ = 78 GPa, $E_{Pt}$ = 168 GPa) the tip material contribution to effective Young's modulus can be significant, particularly for gold-coated cantilevers, resulting in effective increase of contact area.

The contact radius, $a$, is related to the indentation depth as $a = \sqrt{w_0 R}$, or

$$a = \left( \frac{3P}{4E^*} \right)^{\frac{1}{3}} R^{\frac{1}{3}} \tag{5}$$

In PFM imaging, the load acting on the tip $P = k\, d_0$ is exerted by the cantilever having spring constant $k$ at setpoint deflection $d_0$. For typical imaging conditions, the setpoint deflection is ~100 nm, and the spring constant of the cantilever $k$ varies from ~0.01 to ~100 N/m. Consequently, imaging can be done under a range of loads spanning at least 4 orders of magnitude from 1 nN to 10 μN. Note that the contact area is only weakly dependent on effective Young's modulus, which changes by no more than ~ 50% for different tip-surface material pairs, thus resulting only in minor deviations from rigid tip-elastic plane behavior analyzed below. From Eq. (5), the ratio $\eta$ of the tip radius to the contact radius, as a function of tip radius for different loads, is shown in Fig. 2b. Shown for comparison are critical ratios for $\kappa = 100$ and $\kappa = 300$. For small indentation forces (10 nN), the capacitive contribution from the spherical part of the tip dominates for tip radius $R > 10$ nm. However, for large indentation forces commonly used in PFM, for typical tip radii of the order of 50-100 nm, the capacitive contribution from the contact area dominates.

Furthermore, tip flattening due to wear (inevitable under these conditions) and elastic deformation of the tip material will further increase the contact radius. Thus, applicability of the electrostatic sphere-plane model Eq. (1) to the description of the fields inside ferroelectric material to small indentation forces and large tip radii is limited. In addition to the theoretical arguments developed above, strong experimental evidence towards the validity of the analysis above is that the resolution in PFM experiments can be as high as ~ 5 nm using metal coated probes with typical radius of curvature of order of 50-70 nm, which clearly indicates dominant contribution of contact area to the measured PFM signal.

Thus, for quantitative description of the fields in a ferroelectric material required for the analysis of the PFM spectroscopy and domain patterning processes under realistic conditions, contributions from both the spherical part of the tip and the contact area must be taken into account depending on imaging conditions.

## IV. Fields in the Weak Indentation Limit

In the weak indentation limit, contact area contribution to the tip surface interactions can be ignored ($C_{ca} << C_d$ in Fig. 1 c), and the field distribution in the tip-surface junction and inside the ferroelectric material can be analyzed using a purely electrostatic sphere plane model ignoring the mechanical effect of the tip and the electroelastic coupling in the material. To estimate the electrostatic potential distribution inside anisotropic ferroelectric material in the rigid dielectric limit, we use the image charge method.[44,45] The image charge distribution in the tip can be represented by the set of image charges $Q_i$ located at distances $r_i$ from the center of the sphere such that:

$$Q_{i+1} = \frac{\kappa - 1}{\kappa + 1} \frac{R}{2(R + d) - r_i} Q_i, \qquad r_{i+1} = \frac{R^2}{2(R + d) - r_i}, \tag{6a,b}$$

where $R$ is the tip radius, $d$ is the tip-surface separation, $Q_0 = 4\pi\varepsilon_0 RV$, $r_0 = 0$ and $V$ is the tip bias. The tip-surface capacitance is $C_d(d,\kappa)V = \sum_{i=0}^{\infty} Q_i$ and for the conductive tip-dielectric surface

$$C_d = 4\pi\varepsilon_0 R \sinh\beta_0 \sum_{n=1} \left(\frac{\kappa-1}{\kappa+1}\right)^{n-1} \left(\sinh n\beta_0\right)^{-1}, \tag{7}$$

where $\beta_0 = \mathrm{arccosh}\left((R+d)/R\right)$. In the limit of small tip-surface separation, $C_d$ converges to the universal "dielectric" limit, Eq. (1).[41,42] For conductive surfaces, $\kappa \to \infty$, capacitance diverges logarithmically. Potential and field distributions inside the dielectric material can be found using a modified image-charge model as described by Mele:[46]

$$V_i(\rho, z) = \frac{Q_i}{2\pi\varepsilon_0(\kappa+1)} \frac{1}{\sqrt{\rho^2 + (r_i + z/\gamma - d - R)^2}}, \tag{8}$$

where $\gamma = \sqrt{\kappa_{33}/\kappa_{11}}$ and $\rho$ is radial coordinate along the surface. The total potential inside ferroelectric in the image-charge model is

$$V_{ic}(\rho, z) = \sum_{i=0}^{\infty} V_i . \tag{9}$$

Far from the contact area, $\rho, z >> R$, the potential distribution is similar to that generated by a point charge $Q = C_d V$ on the anisotropic dielectric surface:

$$V_{ic}(\rho, z) = \frac{C_d V}{2\pi\varepsilon_0(\kappa+1)} \frac{1}{\sqrt{\rho^2 + (z/\gamma)^2}} . \tag{10}$$

A similar approximation was used in Ref. [38] to describe the domain switching processes for the domain size larger than the tip radius. For small separations from the contact area, the point-charge approximation is no longer valid and a full description using Eqs. (8,9) is required. A simplified description of the fields inside the material far from the tip-surface junction can still be obtained using an image-charge model of charge $Q = C_d V$ located at distance $h$ *above* the surface, where $h$ is suitably chosen parameter. Simple analysis of Eqs. (6a,b) indicates that the potential is dominated by the image charges located close to the dielectric surface. This behavior is illustrated in Fig. 3a demonstrating the dielectric constant dependence of tip capacitance and dimensionless charge surface separation, $S(\kappa) = h / R$, defined as the first moment of the image charge distribution

$$S(\kappa) = 1 - \frac{1}{RC_d} \sum_{i=0}^{\infty} Q_i r_i \ . \tag{11}$$

For large dielectric constants, the effective charge-surface separation is much smaller than the tip radius of curvature, reflecting the charge concentration near the tip-surface contact. The potential distribution in the ferroelectric for $\kappa = 100$ calculated for sphere-plane and point charge models for $Q = C_d V$ is illustrated in Fig. 3b. It is clear that, for $z >> R$, the potential distribution follows Eq. (10), for small $z$ the exact form of Eq. (9) must be taken into account to adequately represent the potential distribution directly below the tip. The cross-over from sphere-plane to asymptotic point charge behavior occurs at distances comparable to the tip radius. Given the characteristic size of the tip of order of 10 – 200 nm, a rigorous description of the early stages of polarization switching phenomena in the weak indentation limit necessitates the use of Eq. (9). This is particularly the case for applications such as ultrahigh density ferroelectric recording in thin films, in which minimum achievable domain size (radius ~ 20 nm)[8] is comparable to tip radius of curvature.

Provided the electrical potential distribution below the tip is known (e.g. Eq.(9)) corresponding stress and strain fields can be reconstructed using isotropic Green's function method as suggested by Felten et al.,[47] or calculated numerically using finite elements methods.[48] These approaches provide approximate description of electroelastic field structure in the weak indentation case, for which rigorous analytical solution taking into account electromechanical coupling effects can not be obtained. Furthermore, the numerical analysis of the analytical solutions for the strong indentation case developed below illustrates the validity of these approximate approaches.

## V. Potential and Field in the Strong Indentation Limit

Quantitative imaging of the electromechanical properties of a ferroelectric surface requires good contact between the tip and the surface so that the surface potential on the ferroelectric below the tip is equal to the tip potential, minimizing the dielectric gap effect in the contact area. At the same time, the contribution of the spherical part of the tip not in contact with the surface to the capacitance can usually be neglected ($C_d << \left(C_{ca}^{-1} + C_q^{-1}\right)^{-1}$ in Fig. 1c), as discussed in Section III. In this strong indentation regime, the description of the field distribution below the tip requires both electrostatic and electroelastic coupling effects to be taken into account to adequately describe the PFM imaging and polarization switching mechanisms. The importance of such coupling was demonstrated by Abplanalp for stress-induced high-order switching processes.[16] In this regime, description of the PFM contrast mechanism is similar to the one for the indentation of a piezoelectric material by a biased conductive indentor. Summarized below are the exact results for the full field distributions inside the transversely isotropic piezoelectric half plane subjected to spherical indentation.

### V.1. Existing results on the contact problem for piezoelectric materials

In the last decade, substantial progress, based on advances in potential theory[22,23] has been made in obtaining closed form exact solutions in elementary functions for a number of 3-D cracks and contact problems in transversely isotropic piezoelectric solids.[49,50] These results are relevant for those contact problems that model the PFM imaging mechanism in the strong indentation regime. The following works should be mentioned in this connection. Chen and Ding[51] have derived electroelastic fields for the spherical punch problem; however, their results are given in the form that does not explicitly identify the combinations of electroelastic constants in whose terms the fields are expressed (these combinations are identified in our analysis, see the text to follow). In the work of Giannakopoulos and Suresh[52] and a follow-up work of Giannakopoulos,[53] three punch geometries were considered: spherical, conical, and circular flat. In these works, electroelastic fields in the plane $z=0$ were given in the closed form. For the full fields, integral representations were given (results in this form make it more difficult, as compared with solutions in elementary functions, to distinguish the contributions of the bias- and the stress effects). We also note that their boundary conditions contain an unclear statement that $\sigma_{zz}=0$ at the edge of the contact zone $(\rho=a)$: being correct for the spherical and conical shapes, it is incorrect for the flat punch (moreover, $\rho=a$ is actually a singularity point in this case, as seen from the Table 2 in Ref. [52]).

Karapetian *et al.*[24] has established the general correspondence principle between the elastic and the piezoelectric solutions for transversely isotropic materials, and considered, as an illustration, the problem of a circular flat rigid punch on a piezoelectric half-space, under applied normal force and tilting moment. In the following section, this principle is applied to obtain solutions in elementary functions for the full fields inside the ferroelectric medium for the spherical indentation. These solutions allow the contributions of the bias- and the stress effects to be differentiated. The asymptotic behavior of the fields far from the contact area is also determined.

## V.2. Explicit Solution of the Problem of a Spherical Hertzian Indenter on the Piezoelectric Half-space (Full Fields)

We now consider a transversely isotropic piezoelectric half-space (with the planes of isotropy parallel to the boundary) pressed upon by a spherical Hertzian indenter. Here we utilize the general elastic-piezoelectric correspondence principle (Karapetian *et al.*, 2002)[24] that expresses full piezoelectric fields in terms of the purely elastic ones for the corresponding elasticity problem. The purely elastic result for the stated problem was given by Hanson (1992).[54]

Boundary conditions in the considered piezoelectric problem are as follows. For the vertical displacement $u_z$, electric potential $\psi$, shear stresses $\tau_z \equiv \sigma_{zx} + i\sigma_{yz}$, normal stress $\sigma_{zz}$, and the normal component of the electric displacement $D_z$ in the plane $z = 0$ are $u_z = w(\rho,\phi) = w_0 - \beta\rho^2$ and $\psi = \psi_0$ for $0 \le \rho < a$, $\sigma_{zz} = 0$ and $D_z = 0$ for $\rho > a$, and $\tau_z = 0$ for $0 \le \rho < \infty$, where $a$ is the radius of the contact zone, $w_0$ is the displacement of the rigid sphere and $\rho,\phi$ are polar coordinates. The prescribed vertical displacement of the boundary $w(\rho,\phi)$ is determined by the shape of the indenter. For spherical indentation, $\beta = 1/2R$ where $R$ is the radius of curvature of the tip. The electric potential $\psi_0$ is constant and is determined by the tip potential. From geometry of the configuration, $w_0 = 2a^2\beta$ so that $w(\rho,\phi) = \left(2a^2 - \rho^2\right)/2R$.

Boundary conditions in Hanson's solution are given in terms of the prescribed force on the punch, rather than prescribed displacement. Therefore, we first obtain the piezoelectric solution corresponding to Hanson's solution. Then, we find a solution to the piezoelectric boundary value problem formulated above by using the "stiffness relation" between the pair (displacement $w_0$, electric potential $\psi_0$) and the pair (force, charge).

In the correspondence principle, the piezoelectric analogues of the terms occurring in purely elastic solutions are identified in the "Correspondence Tables" 1 and 2 of the work of Karapetian *et al.*[24] In the context of the punch problem, Table 2 is relevant.

Thus, application of the correspondence principle yields the following solution of the boundary value problem stated above. For convenience, we present it as a superposition of the two sub-problems: (A) the sub-problem with purely mechanical boundary conditions

$$u_z = w(\rho,\phi) = \frac{2a^2 - \rho^2}{2R}, \qquad 0 \le \rho < a \tag{12}$$

$$\sigma_{zz} = 0, \qquad \rho > a \tag{13}$$

$$\tau_z = 0, \qquad 0 \le \rho < \infty \tag{14}$$

and zero electric boundary conditions and (B) the sub-problem with purely electrical boundary conditions

$$\psi = \psi_0 \qquad 0 \le \rho < a \tag{15}$$

$$D_z = 0 \qquad \rho > a \tag{16}$$

and zero mechanical boundary conditions.

The solutions for normal displacement and electric potential of the sub-problem (A) are as follows:

$$u_z = -\frac{H^*}{\pi R}\sum_{j=1}^{3}\frac{m_j^*}{\gamma_j^*}\left(N_j^* C_1^* + L_j^* C_2^*\right)\left[\left(2a^2 + 2z_j^2 - \rho^2\right)\arcsin\left(\frac{l_{1j}}{\rho}\right) + \frac{\left(3l_{1j}^2 - 2a^2\right)\left(l_{2j}^2 - a^2\right)^{1/2}}{a}\right] \tag{17}$$

$$\psi = -\frac{H^*}{\pi R}\sum_{j=1}^{3}\frac{k_j^*}{\gamma_j^*}\left(N_j^* C_1^* + L_j^* C_2^*\right)\left[\left(2a^2 + 2z_j^2 - \rho^2\right)\arcsin\left(\frac{l_{1j}}{\rho}\right) + \frac{\left(3l_{1j}^2 - 2a^2\right)\left(l_{2j}^2 - a^2\right)^{1/2}}{a}\right] \quad (18)$$

where corresponding constants are defined in Appendix A. The solutions of the sub-problem (B) are as follows:

$$u_z = -\frac{2\psi_0 H^*}{\pi}\sum_{j=1}^{3}\frac{m_j^*}{\gamma_j^*}\left(N_j^* C_3^* + L_j^* C_4^*\right)\arcsin\left(\frac{a}{l_{2j}}\right) \quad (19)$$

$$\psi = -\frac{2\psi_0 H^*}{\pi}\sum_{j=1}^{3}\frac{k_j^*}{\gamma_j^*}\left(N_j^* C_3^* + L_j^* C_4^*\right)\arcsin\left(\frac{a}{l_{2j}}\right) \quad (20)$$

where corresponding constants are defined in Appendix A. Full solutions for other field components are presented in Appendix B. Similar solutions for the flat punch and conical indentor corresponding to other limiting cases of tip geometry are published elsewhere.[55]

## VI. Stiffness Relations and Piezoresponse Force Microscopy Mechanism

The theoretical approach outlined in Section V yield full fields under the indentor, expressed in elementary functions. In this section, we analyze stiffness relations for the spherical indentation and the relation to the PFM contact mechanics and imaging mechanism.

### VI.1. Stiffness Relations for Spherical Indentor

The solutions in Section V imply the following stiffness relations that interrelate applied force $P$ and concentrated charge $Q$ (required to maintain prescribed displacement, $w_0$ and potential, $\psi_0$ and obtained by integrating normal stress, $\sigma_{zz}$, and electric displacement, $D_z$, at $z = 0$ over the contact region) to $w_0$ and $\psi_0$. Integration of the stress components $\sigma_{zz} = -2C_1^*/(\pi^2 R)(a^2 - \rho^2)^{1/2}$ in the sub-problem (A) and $\sigma_{zz} = -\psi_0 C_3^*/(\pi^2)(a^2 - \rho^2)^{-1/2}$ in the sub-problem (B) over the contact area yields:

$$P = \frac{4a^3 C_1^*}{3\pi R} + \frac{2a\psi_0 C_3^*}{\pi} = \frac{4aw_0 C_1^*}{3\pi} + \frac{2a\psi_0 C_3^*}{\pi} \tag{21}$$

Similar integration of the electric displacement components $D_z = 2C_2^*/(\pi^2 R)(a^2 - \rho^2)^{1/2}$ in the sub-problem (A) and $D_z = \psi_0 C_4^*/(\pi^2)(a^2 - \rho^2)^{-1/2}$ in the sub-problem (B) over the contact area yields:

$$Q = \frac{4a^3 C_2^*}{3\pi R} + \frac{2a\psi_0 C_4^*}{\pi} = \frac{4aw_0 C_2^*}{3\pi} + \frac{2a\psi_0 C_4^*}{\pi} \tag{22}$$

We further find from the results presented in Appendix A that $C_2^* = -C_3^*$ and therefore the electromechanics of the spherical indentation is described by the following set of equations:

$$P = \frac{4a^3 C_1^*}{3\pi R} + \frac{2a\psi_0 C_3^*}{\pi} \tag{23}$$

$$Q = -\frac{4a^3 C_3^*}{3\pi R} + \frac{2a\psi_0 C_4^*}{\pi} \tag{24}$$

where $w_0 = \frac{a^2}{R}$ and constants $C_1^*$, $C_3^*$, and $C_4^*$ are materials dependent coefficients defined in Appendix A.

Note that the stiffness relations Eqs. (21,22) derived here have the same structure as the ones of Giannakopoulos and Suresh,[52] but contain numerically different constants. We believe that our relations are correct since they have been verified (by rather lengthy calculations) to be in agreement with independently obtained results of Chen and Ding.[51] Moreover, a single indentation piezocoefficient relates the indentation force and potential in Eq. (23) and charge and displacement in Eq. (24), similarly to the direct and inverse piezoelectric effect in the uniform field case.

The structure of Eqs. (23,24) allows a straightforward interpretation by considering physical meaning of the individual terms. From Eq. (23), the relationship between the indentation depth $w_0$ and force for zero tip bias, $\psi_0 = 0$ can be found as $w_0 = \left(3\pi P/4C_1^*\right)^{2/3} R^{-1/3}$. This is equivalent to the classical Hertzian indentation, where the effective Young's modulus is related to the materials constants for anisotropic piezoelectrics as $E^* = C_1^*/\pi$. Thus, constant $C_1^*$ can be identified as *indentation elastic stiffness*. For a large indentor radius of curvature, $R \to \infty$ (flat contact), the second term in Eq. (24) relates indentor charge to the contact area as $Q = 2a\psi_0 C_4^*/\pi$. This can be compared with the capacitance of the disc on the dielectric substrate, $C_{ca} = 4\kappa\varepsilon_0 a$, providing the relationship between $C_4^*$ and the effective dielectric constant, $\kappa_{eff}$, as $\kappa_{eff} = C_4^*/2\pi$. Thus, constant $C_4^*$ can be identified as *indentation dielectric constant*. Finally, the second term in Eq. (23), $P = 2a\psi_0 C_3^*/\pi$, and first term in Eq. (24), $Q = -4a^3 C_3^*/(3\pi R)$, describe the electroelastic coupling in the material and allow the electrical response to the mechanical indentation and mechanical response to indentor bias to be estimated. The constant $C_3^*$ is thus identified as *indentation piezocoefficient*.

Interestingly, electroelastic coupling in the spherical indentation problem closely resembles that for the uniform field. In both cases, the same electromechanical constant describes the coupling between the charge and the force and the displacement and the potential. For the piezoelectric material in a uniform electric field, the deformation is related to the potential as $h = d_{33}\psi_0$ and charge is related to the force as $Q = d_{33}P$ with the same proportionality coefficient, $d_{33}$. In the spherical indentation problem, for the weak electromechanical coupling, the load can be related to the contact radius and penetration depth as $P = 4aw_0 C_1^* / (3\pi)$. The electroelastic coupling term in Eq. (23) then becomes $w_0 = \left(3C_3^* / 2C_1^*\right)\psi_0$. Similarly, the first term in Eq. (24) becomes $Q = -\left(C_3^* / C_1^*\right)P$. Thus, $C_3^* / C_1^*$ is a single piezoelectric constant describing coupling between the force and the charge and the potential and displacement, similarly to the $d_{33}$ in the uniform field case.

To summarize, the stiffness relations for the spherical piezoelectric indentation can be interpreted as a sum of elastic, electroelastic, and electrostatic contributions. Using the analogy with purely elastic and rigid dielectric solutions, corresponding coupling coefficients can be interpreted as the indentation elastic stiffness (analogous to Young's modulus in planar case), indentation piezocoefficient (analogous to $d_{33}$ in planar case), and indentation dielectric constant (analogous to $\varepsilon_{33}$ in planar case) of the material. As in the uniform field case, the same coefficient describes the coupling between the charge and the force and the displacement and the potential, illustrating the similarity between the two geometries. At the same time, the coupling coefficients *per se* are now complex functions of the complete set of electroelastic constants of the material.

### VI.2. Effect of Materials Properties on Coupling Coefficients

The fields given in Section V depend on the material properties in a rather complex way. The stiffness relations that relate the indentation depth, load, and tip bias include three coupling coefficients which are a complex algebraic functions of 9 of the 10 electroelastic constants for transversely isotropic piezoelectric medium. Based on their structure, the coupling coefficients can be interpreted as indentation elastic stiffness, indentation piezocoefficient, and indentation dielectric constant of the material. Additional insight into the mechanism of spherical indentation can be obtained from analysis of relative contributions of different electroelastic constants to the coupling coefficients $C_k^*$.

Electroelastic properties of the solid can be described either in terms of elastic compliances $s_{ij}$ [m$^2$/N], piezoelectric constants $d_{ij}$ [C/N or m/V], and dielectric permittivities $\varepsilon_{ij}$ [F/m], or in terms of elastic stiffness constants $c_{ij}$ [N/m$^2$], piezoelectric constants $e_{ij}$ [C/m$^2$ or Vm/N], and dielectric permittivities $\varepsilon_{ij}$ [F/m]. These sets of constants are interrelated through the following relationships in the Voigt notation: $d_{nj} = e_{ni} s_{ij}$, $e_{nj} = d_{ni} c_{ij}$, $s_{ij} = c_{ij}^{-1}$, and $c_{ij} = s_{ij}^{-1}$. In order to clarify the relative contributions of different electroelastic constants to coupling coefficients, a sensitivity function of the coupling coefficient, $C_k^*$, is defined as the logarithmic derivative of $C_k^*$ with respect to selected electroelastic constant $f_{ij}$, $S_k(f_{ij}) = \delta \ln C_k^* / \delta \ln f_{ij}$. Numerically, the sensitivity function is calculated as

$$S_k(f_{ij}) = \frac{C_k^*(f_{ij} = 1.01 f_{ij}^0) - C_k^*(f_{ij} = 0.99 f_{ij}^0)}{0.02 C_k^*(f_{ij} = f_{ij}^0)} \tag{27}$$

where $f_{ij}$ is a selected electroelastic constant and $f_{ij}^0$ is a reference value for that constant. A positive value of $S_k(f_{ij})$ implies that a higher constant value favors coupling, while for negative values of $S_k(f_{ij})$, the coupling coefficient decreases with the constant. $S_k(f_{ij}) \approx 0$ indicates that the coupling coefficient is independent of that property. Sensitivity of coupling coefficients $C_k^*$ for polycrystalline PZT6b in the ($c_{ij}$, $e_{ij}$, $\varepsilon_{ij}$) and ($s_{ij}$, $d_{ij}$, $\varepsilon_{ij}$) representations is shown in Fig. 5. The indentation elastic stiffness, $C_1^*$, is dominated by the elastic stiffnesses, $c_{ij}$, the dominant contribution coming from the $c_{33}$ and $c_{44}$. In the ($s_{ij}$, $d_{ij}$, $\varepsilon_{ij}$) representation, $C_1^*$ decreases for high elastic compliances, the dominant contribution coming from $s_{33}$ and $s_{44}$. $C_1^*$ only weakly depends on piezoelectric and dielectric constants, in accordance with the analogy of $C_1^*$ with the effective Young's modulus for planar case.

The indentation piezocoefficient, $C_3^*$, is determined primarily by the piezoelectric constants $e_{33}$ and $e_{15}$ and dielectric constants $\varepsilon_{11}$ and $\varepsilon_{33}$, while it is virtually insensitive to $e_{33}$ and elastic stiffnesses, $c_{ij}$. The indentation piezocoefficient increases with $e_{33}$, $e_{15}$, and $\varepsilon_{33}$ and decreases with $\varepsilon_{11}$. This can be understood from the analysis of the field distribution below the indentor. For large $\varepsilon_{33}$, the potential is concentrated below the tip along the $z$-axis, thus increasing the electromechanical coupling, while for large $\varepsilon_{11}$ lateral spreading of the field reduces the coupling coefficient. In the ($s_{ij}$, $d_{ij}$, $\varepsilon_{ij}$) representation, the contributions of all electroelastic constants to $C_3^*$ are comparable.

The indentation dielectric constant, $C_4^*$, is determined primarily by dielectric constants $\varepsilon_{11}$ and $\varepsilon_{33}$, with other elastic and piezoelectric constants providing only minor contributions. This can be understood from the comparison with the rigid dielectric problem, for which the effective dielectric constant for the point charge is the geometric average of the principal values of dielectric constant tensor, $\varepsilon_{eff} = \sqrt{\varepsilon_{11}\varepsilon_{33}}$.

In order to obtain further insight into relative contribution of elastic, dielectric and piezoelectric constants of material into the coupling coefficients in Eqs. (23, 24), the scaling analysis of elastic, piezoelectric and dielectric contributions was performed using formulae in Appendix A. To estimate the contribution of piezoelectric constants $e_{ij}$ to the indentation elastic stiffness $C_1^*$, that latter was calculated as a function of parameter $\gamma$ for material with fictitious set of electroelastic constants ($c_{ij}$,,$\gamma$ $e_{ij}$, $\varepsilon_{ij}$), where the original set of electroelastic constants ($c_{ij}$, $e_{ij}$, $\varepsilon_{ij}$) corresponds to $BaTiO_3$ and $LiNbO_3$, as illustrated in Fig. 6 a. Note that while $\gamma$ can be arbitrarily small corresponding to zero electromechanical coupling, for real material the value of $\gamma$ is limited. Fig. 6a illustrates that for $\gamma = 1$ (real material) contribution of piezoelectric coupling to overall elastic properties is ~ 10 % for $BaTiO_3$ and 6.5 % for $LiNbO_3$. Interestingly, both for $\gamma \to 0$ and for $\gamma \to \infty$ indentation elastic stiffness adopts the finite value; however, this value is determined by different combination of elastic constants of materials. Similar analysis can be performed for indentation piezocoefficient and indentation dielectric constant, as illustrated in Fig. 6 b,c. As expected, indentation piezocoefficient is almost linear in $\gamma$. At the same time, indentation dielectric constant is virtually independent on $\gamma$ for $\gamma \to 0$ and is determined solely by $\varepsilon_{ij}$ in this limit. For $\gamma \to \infty$ indentation dielectric constant is determined primarily by $e_{ij}$ and diverges as $C_4^* \sim \gamma^2$. For $\gamma = 1$, the contribution of piezoelectric coupling to overall dielectric properties is ~ 24.6 % for $BaTiO_3$ and 11.4 % for $LiNbO_3$. This scaling analysis allows the behavior of other characteristic properties to be predicted. For example, the maximal electrostatic potential in the material in mechanical problem is linear in indentation piezocoefficient and inversely proportional to indentation dielectric constant, $\psi_{max} \sim hC_3^*/C_4^*$. From Fig. 6a,b,c, $\psi_{\max} \sim \gamma$ for $\gamma \to 0$ and $\psi_{\max} \sim \gamma^{-1}$ for $\gamma \to \infty$, as illustrated in Fig. 6d. This behavior is counterintuitive, since simple analysis predicts that potential generated in the material will increase linearly with elecromechanical coupling. Note that the maximum potential that develops inside the material during the indentation is limited and "conventional" materials such as $BaTiO_3$ and $LiNbO_3$ correspond to nearly optimal values of coupling coefficients.

Similar analysis can be performed for indentation elastic stiffness, indentation piezocoefficient and indentation dielectric constant by scaling elastic, ($\gamma\, c_{ij}$, $e_{ij}$, $\varepsilon_{ij}$), and dielectric, ($c_{ij}$, $e_{ij}$, $\gamma\, \varepsilon_{ij}$), properties, as illustrated in Fig. 7 a,b,c. As expected, indentation elastic stiffness scales linearly with elastic constants and is only weakly dependent on dielectric constants. Indentation piezocoefficient adopts finite limiting values both for $\gamma \to 0$ and for $\gamma \to \infty$. Finally, indentation dielectric constant shows non-trivial scaling as $C_4^* \sim \gamma^{-1}$ for $\gamma \to 0$ and $C_4^* \sim const$ for $\gamma \to \infty$ for mechanical case and $C_4^* \sim const$ for $\gamma \to 0$ and $C_4^* \sim \gamma$ for $\gamma \to \infty$ for dielectric case. The resulting behavior of maximum potential inside the material is illustrated in Fig. 7d. The scaling behavior for effective indentation properties is summarized in Table 2.

### VI.3. Effective Piezoresponse Amplitude and Dielectric Constant

The stiffness relations relating the indentation depth, indentation force, and indentor potential can be immediately used for the description of the PFM imaging mechanism, and the determination of the relative contribution of elastic, electrostatic, and electroelastic coupling terms to the tip potential and force.

The relative contributions of indentor potential and penetration depth to force and charge can be determined from the stiffness equations. From Eq. (23) for small tip potential, the force is primarily determined by the indentation depth (elastic term dominates), while for large tip potentials, the electroelastic contribution to the force is larger. The boundary between the two regimes is given by $\psi_0 = 2C_1^*/(3C_3^*)w_0$. Similarly, from Eq. (24) for small tip potentials, the charge is dominated by the electroelastic coupling, while for larger tip potentials the charge is determined by the electrostatic properties of the tip-surface junctions. The boundary between the two regimes is $\psi_0 = -2C_3^*/(3C_4^*)w_0$. From the magnitudes of the coupling coefficients for ferroelectric materials (Table 1) calculated using formulae in Appendix A, the ratio $C_1^*/C_3^*$ is typically two orders of magnitude larger than $C_3^*/C_4^*$, giving rise to the plot in Fig. 8a. In region I for small tip biases, the force is dominated by the penetration depth (elastic coupling), and the tip charge is determined by electroelastic coupling. In this case, the elastic response of material (PFM signal) can be controlled by the load applied by the cantilever; however, the electrical field (e.g., relevant to the polarization switching processes) is primarily determined by the force (electromechanical coupling), rather then tip bias. In region II for moderate tip biases, the force is still dominated by the penetration depth (elastic), but the tip charge is now determined primarily by the electrostatic term. In this case, both elastic response of material and electrical field distribution can be controlled independently by the applied load and bias. Finally, in region III for large biases the penetration depth is determined by the electroelastic term. In this case, the electrical field distribution can be controlled by applied tip bias; however, the electroelastic contribution to the stress and strain field dominates and the latter can not be controlled independently by applied load.

The plot in Fig. 6a allows the dominant coupling mechanism to be related to the experimental conditions. Experimentally accessible are the indentation force and tip bias, rather than the indentation depth, and the correspondence between regimes in Fig. 8a and experimental conditions can be established using stiffness relations Eq. (23), as illustrated in

Fig. 8b. In a realistic PFM experiment the contact force is limited by the capacitive tip-surface interaction and capillary force as $P > F_{cap} + F_c = CV^2 + F_c$, limiting the range of accessible bias-indentation phase space.[56] Nevertheless, the plot in Figs. 8a,b illustrates the relative contribution of electrostatic, electroelastic and elastic components to the electrical and mechanical characteristics of the tip and electroelastic fields in the material in PFM.

The relationship between indentation depth and tip bias for a given elastic force required for the description of the PFM imaging mechanism can be found from the stiffness relation Eq. (23). This solutions for PZT6b for a tip radius of curvature $R$ = 50 nm and several indentation forces are illustrated in Fig. 9a,b. Eq. (23) generally has one solution for positive biases and one or three solutions for negative biases. In the latter case, two emerging solutions correspond to negative contact areas and are physically meaningless. Note that for small indentation forces and large positive biases the penetration depth, contact area, and charge become effectively zero, since the indentor is effectively pushed out from the material because of the inverse piezoelectric effect. For large indentation forces and small biases, the elastic contribution to the indentation depth dominates and the indentation depth is linear in tip bias.

For small modulation amplitudes, the PFM amplitude is $A_{piezo} = dw_0/d\psi_0$, where the functional dependence of $w_0$ on the bias is given by Eq. (23). Shown in Fig. 9c is the bias dependence of the piezoresponse amplitude for polycrystalline PZT6b calculated for a tip radius $R$ = 50 nm for different indentation forces. For small indentation forces, the response amplitude is zero for large biases. This corresponds to the zero indentation $w_0$, in which case the electromechanical response of the material effectively prevents the penetration of the tip. Note that in this case the description of PFM mechanism requires taking into account the electrical field produced by the spherical part of the tip not in contact with the surface (cross-over to weak indentation), as analyzed by Felten et al.[47] For large indentation forces, contact geometry is only weakly affected by the electromechanical response. In this case, where the dominant contribution to the load is mechanical ($P_{mech} > P_{piezo}$), the indentation depth is related to tip bias as $w_0(V) \approx w_0 - V C_3/C_1$. Hence, the effective electromechanical response measured by PFM is $A_{piezo} = C_3/C_1$, in agreement with phenomenological arguments developed in Section VI.1. A similar analysis can be performed for the effective dielectric constant defined as $\kappa_{eff} = Q/(4a\varepsilon_0)$, where $a$ is the contact radius. Bias dependence of the effective dielectric constant is illustrated in Fig. 9d. The indentation dielectric constant in this case is bias dependent due to the change in contact radius and relative contribution of electromechanical coupling to dielectric properties. Noteworthy, the bias dependencies of $A_{piezo}$ and $\kappa_{eff}$ are functionally identical, stemming from the structure of the stiffness relations.

To determine the contribution of different electroelastic constants to the piezoresponse amplitude, the sensitivity function for piezoresponse amplitude, $A_{piezo}$, was calculated as shown in Fig. 10. In the ($c_{ij}$, $e_{ij}$, $\varepsilon_{ij}$) representation $A_{piezo}$ is dominated by the elastic stiffnesses $c_{33}$ and $c_{44}$, piezoelectric constants $e_{33}$, $e_{15}$, and dielectric constants $\varepsilon_{11}$ and $\varepsilon_{33}$. Piezoresponse decreases with elastic stiffnesses and increases with piezoelectric constants, as expected. In the ($s_{ij}$, $d_{ij}$, $\varepsilon_{ij}$) representation, piezoresponse is clearly dominated by the piezoelectric constants $d_{33}$ and $d_{15}$ and only weakly depends on elastic compliances. Similarly to the indentation piezocoefficient, $C_3^*$, the piezoresponse amplitude increases with $\varepsilon_{33}$ and decreases with $\varepsilon_{11}$. This behavior can be readily understood from the schematics in Fig. 10. Normal component of electric field is related to the vertical strain component by piezoelectric constant $d_{33}$, as shown in Fig. 10c. In the spherical indentation geometry, an additional contribution to response amplitude originates from the lateral component of electric fields related to the vertical strain component by piezoelectric constant $d_{15}$ as shown in Fig. 10d. The ratio between the lateral and vertical field components is determined by $\varepsilon_{11}/\varepsilon_{33}$, thus rationalizing the dominant contributions of these constants to the sensitivity function for $A_{piezo}$.

To establish the correlation between the measured piezoresponse and $d_{33}$ of the material, the calculated piezoresponse coefficient is compared with the piezoelectric constant for a set of polycrystalline lead zirconate-titanate (PZT) materials and several single-crystal ferroelectric materials as shown in Fig. 11a. The numerical values for the corresponding electroelastic constants are obtained from Refs. [57,58,59]. Note that for the polycrystalline PZT materials effective piezoresponse is almost a linear function of $d_{33}$, $A_{piezo} \approx d_{33}$. At the same time, for single-crystal materials such as $BaTiO_3$, $LiNbO_3$ and $LiTaO_3$ the piezoresponse amplitude significantly differs from $d_{33}$. This can be readily understood from the fact that the sensitivity function for $A_{piezo}$ shown in Fig. 10 is strongly affected by $d_{15}$ and dielectric constants. For the single crystalline ferroelectrics, strong anisotropy of piezoelectric and dielectric tensors results in a nontrivial relationship between $A_{piezo}$ and $d_{33}$. In comparison, in polycrystalline materials the dielectric and piezoelectric tensors are more symmetric due to the averaging between the grains with different crystallographic orientation, resulting in good correlation between piezoelectric responses in the spherical and planar geometries. It must be noted that while a linear relationship between $d_{33}$ and $A_{piezo}$ for polycrystalline materials applies for the macroscopic indentation, in which the contact radius is larger than the average grain size, in the typical PFM experiment the small contact area implies that the indentation is performed within a single crystalline grain. Therefore, in general, a quantitative description of the PFM imaging mechanism requires the effective piezoresponse amplitude for spherical indentation to be calculated using the exact formulae in Section V.

A similar analysis can be performed for the electrostatic field distribution below the indentor and is required for the description of bias-induced phenomena in ferroelectric materials. As can be expected from the geometry of the problem, for large separations from the contact area, the potential distribution is reduced to that produced by a point charge. For weak electromechanical coupling (regions II and III in Fig. 8), the indentor charge is determined by the capacitance of the contact area, i.e., coefficient $C_4^*$ in the stiffness relations Eq. (24). Moreover, it can be expected that even for the deviations of the contact geometry from spherical, $C_4^*$ will describe the capacitive contribution to the effective tip charge provided that the contact area is known.[60] From the sensitivity function in Fig. 5, the indentation dielectric constant, $C_4^*$, is determined primarily by dielectric constants $\varepsilon_{11}$ and $\varepsilon_{33}$. This is in agreement with the expected behavior in the rigid electrostatic problem, in which the dielectric response to the point charge is described by the effective dielectric constant $\varepsilon = \sqrt{\varepsilon_{11}\varepsilon_{33}}$, sensitivity function for which, $S_{\varepsilon_{11}}\left(\sqrt{\varepsilon_{11}\varepsilon_{33}}\right) = 1/2$, is consistent with Fig. 5. Illustrated in Fig. 11b is the correlation between $C_4^*$ and $\sqrt{\varepsilon_{11}\varepsilon_{33}}$. Note that both for polycrystalline and single-crystal materials the dielectric properties are described by a linear relationship $C_4^*/2\pi = (1.203 \pm 0.02)\sqrt{\varepsilon_{11}\varepsilon_{33}}$. This analysis, combined with the scaling analysis in Section VI.2, illustrates that the contribution of electromechanical constants to the dielectric properties of the system is of order of 10-20%, thus providing an estimate of the relative error in the analyses of PFM contrast using Green's functions and FEA methods coupled with rigid dielectric solution for electrostatic field in the material.[47,48]

## VII. Structures of the Field.

The solutions for the piezoelectric indentation problem given in Section V provide explicit expressions for elastic and electrical fields inside the material. Because of the linearity of the solution, relative contributions of the mechanical and electrical indentation can be considered, allowing separating force- and bias-induced phenomena in PFM. Analysis of the tip-induced switching phenomena in PFM requires knowledge of the field distributions both below the tip and at the large separation from the contact zone. From the geometry of the problem, it can be expected that for large separations from the contact area asymptotic field behavior can be reduced to the point-charge model and the relevant parameters and applicability limits are determined. At the same time, description of the early stages of the domain nucleation process in which domain size is smaller than the contact radius requires field distributions directly below the tip, since the use of the point charge approximation in this case will result in the unphysical singularities in the field distribution.

Strain and potential distributions below the tip for tip radius $R$ = 50 nm, contact area $a$ = 3 nm and tip bias $\psi_0$ = 1 V (corresponding to indentation force of ~ 100 nN, depending on materials system) for $BaTiO_3$ and $LiNbO_3$ are illustrated in Figs. 12 and 13. Note that for the sub-problem (B) with purely electrical boundary conditions, the potential attains maximum value immediately below the tip and slowly decays for large tip surface separation. The shape of the potential distribution is determined primarily by the anisotropy of the dielectric constant tensor, as can be clearly seen from the comparison of the potential distributions for $BaTiO_3$ and $LiNbO_3$. In comparison, the potential in the sub-problem (A) with purely mechanical boundary conditions is zero directly below the tip and attains maximum value at a certain depth. The maximum potential value in this case is determined by the strength of the electromechanical coupling in the material, as discussed in Section VI.2. Strain distribution below the tip in the sub-problem (A) is maximum for $\rho = 0$ and $z = 0$ and decreases with radial and normal distances, as expected for the spherical indentor geometry. Strain distribution in the sub-problem (B) is zero at the surface due to the choice of boundary conditions and attains maximum value in the material. Note that close similarity exists between the shapes of strain distribution in the sub-problem (B) and of potential distribution in the sub-problem (A).

The normal stress, $\sigma_{zz}$, distribution below the indentor for the sub-problem (B) has a well-known square root singularity at the perimeter of the contact area. Because of the stress contribution to the displacement, a similar singularity exists for the normal component of the displacement vector, $D_z$. At the same time, in the sub-problem (A) there is no singularity at the circumference and both stress and displacement attain maximum value below the tip ($\rho = 0$ and $z = 0$) and decay rapidly with separation from contact area.

The field behavior as a function of depth for several ferroelectric materials is illustrated in Figs. 14 and 15. This behavior is consistent with the 2D plots illustrated in Figs. 12 and 13. Note that the potential below the tip decays much faster for $BaTiO_3$ than for other more uniform ferroelectrics, resulting in the smaller probing depth in the PFM experiment (Fig. 14a). At the same time, the strain distribution below the tip is relatively insensitive to the materials system, since it is determined primarily by the anisotropy of the elastic stiffness tensor $c_{ij}$ (Fig. 14b). Note that the solutions presented in Figs. 14 and 15 correspond to the defined strain boundary conditions, and the difference in materials properties will be reflected in the difference in the indentation force required to achieve this level of indentation. Potential distribution in the sub-problem (A) and strain distribution in the sub-problem (B) are shown in Fig. 14c,d. For the chosen experimental conditions, the electromechanical fields below the tip are dominated by the direct contributions from the tip bias and load, the terms due to the electromechanical coupling being significantly smaller (region II in Fig. 8). However, electromechanical coupling effects are linear in the tip bias and indentation depth. Therefore, relatively small changes in the experimental conditions (particularly tip bias) can change the field distributions so that coupling terms will dominate the direct contributions (strong coupling). It should also be noted that the relevant length scale that determines spatial extent of the electromechanical fields inside the material is the contact radius, related to the tip radius of curvature and indentation force through stiffness relation Eq. (21).

The field behavior as a function of radial coordinate for several ferroelectric materials is illustrated in Figs. 16 and 17. Shown in Fig. 16a is the potential distribution in the electrical problem. Note that this distribution is material independent, in agreement with presented in Table 4. Shown in Fig. 16b is lateral displacement which adopts maximum value at the edge of contact area. By definition, the normal displacement is zero. Both normal stress and electric displacement have square root singularities at the edge of contact area, as shown in Fig. 16c,d. Corresponding behavior for sub-problem (A) is illustrated in Fig. 17. Note that similarly to potential in sub-problem (B), normal displacement in sub-problem (A) is materials independent (Fig. 17a). Corresponding behavior for lateral displacement is illustrated in Fig. 17b. Both normal stress and electric displacement are continuous at the edge of contact area and identically zero outside contact area, as illustrated in Fig. 17c,d.

A prominent feature of the field distributions in Figs. 14, 15 is that nontrivial behavior persists on a length scale comparable to the contact radius. For the distances larger than contact radius, $z > a$, the field distribution quickly adopts the corresponding asymptotic power law behavior. Asymptotic behavior of relevant field quantities in radial and normal directions and its dependence on indentation parameters is summarized in Tables 3 and 4. Note that for both the sub-problem (A) and sub-problem (B) strain and potential decay as $1/z$ similar to the point charge case. The magnitude of the charge due to the mechanical contribution is cubic in the contact radius. For the weak electromechanical coupling, it implies that it is a linear function of indentation force and tip radius. In comparison, charge magnitude due to the tip potential is a linear function of bias and contact radius, making it a much weaker $x^{1/3}$ function of the load and tip radius. This simplified analysis clearly predicts the dominant trends in the PFM experiment for varying experimental conditions such as load and indentor bias.

Interestingly, the crossover to the power law behavior can occur at distances much smaller than the indentation radius. For example, the electric potential due to tip bias adopts a $1/z$ distance dependence at separations as small as ~ 0.3 $a$. This implies that for separations from the indentation zone exceeding the contact radius, the indentor can be modeled with a very good accuracy as a point charge or point force, considerably simplifying the description of the bias- and stress-induced phenomena. Relevant parameters such as force and charge magnitudes including elastic, electric and electromechanical coupling effects can be determined using stiffness relations Eq. (23,24). This behavior is illustrated in Fig. 18a representing a 2D plot of the ratio of the potential distribution below the indentor for $BaTiO_3$ calculated using the exact solution Eq. (20) and rigid dielectric solution Eq. (8) with the point charge magnitude calculated from the stiffness relation Eq. (24). Potential distributions below the tip differ by less than 50% for separations from the tip-surface junction smaller than the contact area. Similar behavior for the ratio of displacement field, $u_z$, calculated from the exact Eq. (17) and from the Green's function for the point force for transversely isotropic material[22] is illustrated in Fig. 18b. As for the electrostatic field, the ratio between the point force and exact solution approaches a value close to unity for very small separations from contact. Note that in both cases the exact asymptotic value of the ratio between point charge/force and exact solutions differs from unity and depends on direction, reflecting the difference in the anisotropy of materials properties in purely elastic, rigid electrostatic, and coupled electroelastic models. Despite this fact, point charge solutions clearly provide a very good approximation for the description of field structure for separations from the contact larger than the contact radius.

This behavior significantly simplifies the description of the PFM mechanism for more complex systems. For example, a good approximation for field structure below the tip in the thin film, as opposed to bulk, ferroelectrics can be achieved using independent image charge and image force models for the electrostatic and elastic field components, provided that the film thickness is larger than the contact radius. The parameters of the corresponding image charge and image forces are then determined by the electrostatic and electric properties of the substrate. However, despite the fact that point charge/force approximation provides a good approximation for field structure even for the small separations from contact, in certain cases the adequate description of the PFM phenomena requires exact structure of the fields taken into account, as illustrated for the examples of PFM signal generation volume and ferroelectric, ferroelastic and high order ferroic switching below.

## VIII. Discussion

Exact closed form results in elementary functions for the full fields in the problem of spherical indentation of piezoelectric material are given in Section V. The full fields under the indentor are linear superposition of solutions for the sub-problem (A) and sub-problem (B). As applied to PFM, this allows the relative contribution of bias and indentation force induced effects on imaging and polarization switching to be separated. It is shown that field distributions have the asymptotic power law form for relatively small separations from the contact area, which in many cases is significantly smaller than the contact radius *per se*.

We now briefly discuss the applicability of obtained solutions for the electroelastic field structure for the description of signal generation volume in PFM and its implications for the polarization switching behavior in ferroelectrics.

### VIII.1 Signal generation volume

The field structure calculated in Section V allows the signal generation volume, and hence the resolution, to be determined. For low modulation frequencies when the tip inertial effects are minimal, the signal generation volume in PFM is given by the field $\partial u_z / \partial \psi$ for $P = const$. Note that the normal displacement in sub-problem (B), which can intuitively be expected to provide the generation volume in PFM, is identically zero at the surface. At the same time, the displacement field in the mechanical problem is tip-bias independent. Thus, the signal generation volume is given by a non-trivial combination of electroelastic fields shown in Figs. 12,13.

To calculate the generation volume, the total displacement field below the tip can be represented as $u_z = u_{z,m}(a) + \psi u_{z,e}(a)$, where the strain field distributions in the sub-problem (A) and sub-problem (B) depend on the contact radius, $a$, and the indentor potential, $\psi$. During imaging, the indentation force $P = const$ and from stiffness relation Eq. (23) the change in tip potential, $\psi = \psi_0 + \delta\psi$, results in the change of contact area, $a = a_0 + \delta a$, as

$$\delta a = \delta\psi \left( \frac{2a_0}{R} \frac{C_1^*}{C_3^*} + \frac{\psi_0}{a_0} \right)^{-1}, \tag{28}$$

The signal generation volume is given by the change in the strain field distribution as

$$SV = \frac{\delta u_z}{\delta \psi} = \frac{u_z(a_0 + \delta a, \psi_0 + \delta\psi) - u_z(a_0, \psi_0)}{\delta\psi} \tag{29}$$

Thus, the signal generation volume in PFM is determined by the combination of the normal displacement fields in the sub-problems with electrical and mechanical boundary conditions. Signal generation volumes for $BaTiO_3$ and $LiNbO_3$ for $R$ = 50 nm, $a$ = 3 nm and $\psi_0$ = 0 V is illustrated in Fig. 19 a,b. In agreement with theoretical expectations, the response is maximal directly below the tip and decays rapidly outside the contact area.

The effective size of the signal generation volume and, thus, the spatial resolution, are controlled by the contact radius, $a$, which is the only relevant parameter in the indentation problem. This suggests that the optimal resolution in PFM can be obtained for small contact areas and moderate indentation forces necessary to prevent tip flattening during imaging. At the same time, the ultimate limit on the PFM resolution is imposed by the electrostatic field contribution from the spherical part of the tip that, in the case of small contact area, will dominate the contact contribution, resulting in loss of resolution (Fig. 1 d) due to cross-over to the weak indentation regime analyzed by Felten et al.[47]

### VIII.2 Implications for PFM polarization switching

Although a rigorous analysis of switching phenomena is an independent problem beyond the scope of this work, we discuss here the applicability of point charge approximation and delineate the cases in which the exact structure of the field is required to analyze the switching dynamics. The knowledge of all components of the electroelastic field distribution under the tip derived in Section V allows direct calculation of the free energy for the switching process. The free energy density contains contributions from several coupling terms:[16,40]

$$\Delta g_{bulk} = -\Delta P_i E_i - \Delta x_\mu X_\mu - \frac{1}{2}\Delta\varepsilon_{ij} E_i E_j - \frac{1}{2}\Delta s_{\mu\nu} X_\mu X_\nu - \Delta d_{i\mu} E_i X_\mu, \quad (30)$$

where the individual terms describe ferroelectric, ferroelastic, ferrobielectric, ferrobielastic and ferroelastoelectric switching respectively, $P_i$ is polarization, $E_i$ is electric field, $x_\mu$ is strain and $X_\mu$ is stress, $i,j$ = 1,2,3, and $\mu$, $\nu$ = 1,..,6. The free energy of the nucleating domain is:

$$\Delta G = \Delta G_{bulk} + \Delta G_{wall} + \Delta G_{dep}, \quad (31)$$

where the first term is the volume change in free energy, the second term is the domain wall energy, and the third term is the depolarization field energy. Using the Landauer model, the domain shape is represented as half ellipsoid with the small and large axis equal to $r_d$ and $l_d$ correspondingly.[61] The domain wall contribution to the free energy in this geometry is $\Delta G_{wall} = b r_d l_d$, where $b = \sigma_{wall}\pi^2/2$ and $\sigma_{wall}$ is the (direction independent) domain wall energy. The depolarization energy contribution depends on the electrostatic conditions on the top surface and for the ferroelectric surface with unscreened polarization charge can be calculated as $\Delta G_{dep} = c r_d^4 / l_d$, where

$$c = \frac{4\pi P_s^2}{3\varepsilon_{11}}\left[\ln\left(\frac{2l_d}{r_d}\sqrt{\frac{\varepsilon_{11}}{\varepsilon_{33}}}\right) - 1\right] \quad (32)$$

only weakly depends on the domain geometry. In the uniform field, the bulk contribution to the domain free energy is $\Delta G_{bulk} = 2 P_s E r_d^2 l_d$ and minimization of Eq. (31) with respect to $r_d$ and $l_d$ allows the critical domain size and activation energy for nucleation to be estimated. It was recognized by Abplanalp (Ref.[40]) and later by Molotskii *et al.*, (Refs.[38,39]) that the field distribution below the PFM tip is strongly non-uniform and the bulk contribution to domain free energy is

$$\Delta G_{bulk} = \int_V \Delta g_{bulk}(\vec{r})dV = 2\pi \int_0^{l_d} dz \int_0^{r_m(z)} \Delta g_{bulk}(r,z) r dr, \qquad (33)$$

where $r_m(z) = r_d \sqrt{1 - z^2/l_d^2}$ is the domain radius at the distance $z$ from the surface. The bulk contribution to the free energy for ferroelectric switching was calculated by Molotskii *et al.* using a phenomenological point charge model, in which the tip is represented by point charge $Q = C_{tip}V$ located at distance $\delta$ from the surface, where $C_{tip}$ is the capacitance between the conductive sphere and anisotropic dielectric half-plane defined in Eq. (1). This corresponds to the weak indentation regime in which the sphere-surface capacitance dominates over the capacitance of the contact area.

It was found that, for domain size $r_d >> \delta$, the critical domain size and the activation energy for nucleation are independent of the effective charge-surface separation $\delta$ and are determined by the materials properties and effective tip charge. This agrees with results of the present work, since, at large separations from the contact area, the potential distribution produced by the tip can be represented by the point charge located on the surface ($\delta = 0$). Therefore, the analysis presented by Molotskii [Ref. 38] becomes rigorous if the tip charge is approximated by the charge at the contact area calculated from the stiffness relations Eq. (23). It should be noted that, for sufficiently high tip bias, switching can be induced by the spherical part of the tip as well, but in this case, rigorous description of the switching process for $l_d$, $r_d < R$ requires calculations of the complete image charge, due to large uncertainties related to choice of the effective charge-surface separation (Section IV).

A similar analysis can be extended to an arbitrary switching mechanism using Eq. (33) to estimate the corresponding free energy. For domain sizes $r_d >> a$ the tip can be modeled as a point charge or point force provided that the singularity in the origin is weak enough to ensure the convergence of the integral in Eq. (33). As summarized in Tables 3 and 4, the asymptotic behavior for potential and strain can be generally represented in the form $f = \left(\rho^2 + (z/\gamma)^2\right)^{-\alpha/2}$, where power α determines the decay rate of the corresponding quantity with the separation from indentation region and γ is the proportionality coefficient reflecting the anisotropy of materials properties. In the rigid dielectric model, $\gamma = \sqrt{\varepsilon_z/\varepsilon_x}$, while in the exact solution in Section V $\gamma_i$, $i = 1, 2, 3$, are the roots of the determinant equation Eq. (A.5) in Appendix A. Corresponding fields are given by the derivatives with respect to the $z$ coordinate, $df/dz$. It can be shown that the bulk contribution to the domain free energy for $l_d >> \gamma\ r_d$ can be calculated as $\Delta G_{bulk} \sim r_d^{2-\alpha}$ for $\alpha < 2$. For bias-induced ferroelectric switching, $\alpha = 1$ and $\Delta G_{bulk} \sim r_d$, in agreement with the analysis in Refs. [38,39].

Similar analysis can be performed for ferroelastic switching, even though in this case the symmetry of the problem requires formation of nontrivial domain structures, for example nucleation of 4 90° domains forming the vortex-type structure required to prevent the formation of energetically unfavorable charged domain walls.[62] At the same time, any high-order ferroelectric switching phenomena including ferrobielectric, ferrobielastic, and ferroelastoelectric are described by the field distributions for which $\alpha \geq 2$ and the integral Eq. (31) does not converge if the asymptotic form of the field is used. This implies that the rigorous description of the high-order ferroelectric switching phenomena requires use of the complete solutions developed in Section V, or suitably chosen extrapolation formulae that adequately represent field distributions at small and large separations from the indentation point, while use of the point charge approximation leads to physically unreasonable divergence of the corresponding free energy. For ferroelectric and ferroelastic switching, the contribution of the volume in the close vicinity of the tip to the free energy can be neglected if $r_d >> a$ and the domain size and activation energy for nucleation become independent of the contact area and are determined solely by the tip charge. This allows arbitrarily large domains to be created for high tip biases. On the contrary, in the higher order ferroelastoelectric, ferrobielectric, and ferrobielastic switching processes, contact contribution to the domain free energy dominates due to the much higher decay rate of the relevant fields. Thus, analyses of the early stages of ferroelectric switching phenomena, as well as higher order ferroic switching, require exact structure of the field to be taken into account and will be reported elsewhere.[63]

## IX. Summary

To achieve quantitative interpretation of PFM, including resolution limits, tip bias- and strain-induced phenomena and spectroscopy, analytical representations for tip-induced electrical and mechanical fields inside the material are derived. The electrostatic potential distribution inside the ferroelectric in the weak indentation limit is obtained using the image charge method. It is shown that, in the general case, this electrostatic solution cannot be reduced to the single point charge approximation, and a complete set of image charges is required to describe switching phenomena. This weak indentation solution implicitly ignores contribution of the tip-surface contact area to the field distributions. At the same time, direct comparison between the sphere-plane capacitance and contact area capacitance estimated using the Hertzian indentation model illustrates that for the typical PFM imaging conditions the contact area contribution to tip-surface capacitance dominates.

These estimates show that rigorous description of the tip-induced phenomena requires solution of the coupled electroelastic problem for spherical indentation of a piezoelectric. Analytical solution of this problem is obtained for the transversely isotropic piezoelectric material using the recently established elastic-piezoelectric correspondence principle. These solutions are used to obtain the electric field and strain distribution inside the ferroelectric material, providing a complete continuum mechanical description of the PFM imaging mechanism for a spherical tip. The relationship between the indentation depth, load, contact area, and indentor bias are given through the stiffness relations that prove to be the extension of Hertzian contact mechanics for a transversely isotropic piezoelectric. The individual coupling coefficients in the stiffness relations can be interpreted as the indentation elastic stiffness, indentation dielectric constant and indentation piezocoefficient, similar to effective Young's modulus, dielectric constant, and piezoelectric constant in the uniform case. Notably, the same piezoelectric coefficient describes charge-force and displacement-bias coupling, demonstrating the similarity between piezoelectric behavior in the spherical and planar geometries. The contributions of different electroelastic constants of the material to the coupling coefficients were investigated.

These rigorous analytical solutions are compared with approximations based on the asymptotic point charge/point force models, and it is shown that crossover to the power law behavior occur at relatively small separations from the contact area. It is also shown that the relevant parameters, including force and charge magnitudes must be obtained from the stiffness relations. Expressions for potential and field in the ferroelectric were used to derive signal generation volume in PFM. The implications for polarization switching phenomena are also analyzed. It is shown that adequate description of late stages of first-order ferroelectric and ferroelastic switching processes can be achieved using the asymptotic representation of the fields; the domain size in this case is determined by the tip charge or force only. At the same time, description of early stages of ferroelectric switching and higher-order switching processes requires detailed description of field distribution below the tip.

**ACKNOWLEDGEMENTS**

Research performed as a Eugene P. Wigner Fellow and staff member at the Oak Ridge National Laboratory, managed by UT-Battelle, LLC, for the U.S. Department of Energy under Contract DE-AC05-00OR22725 (SVK). Discussions with Dr. A. Gruverman (NCSU) and Dr. A.P Baddorf (ORNL) are greatly appreciated.

Table 1. Coupling constants for different materials

| Material | $C_1$, $10^{11}$ N/m$^2$ | $C_3$, N/Vm | $C_4$, $10^{-9}$ C/mV |
|---|---|---|---|
| $BaTiO_3$ | 4.03 | 15.40 | 48.54 |
| $LiNbO_3$ | 6.47 | 7.52 | 3.11 |
| $LiTaO_3$ | 7.80 | 8.80 | 2.81 |
| PZT6B | 3.60 | 25.60 | 23.63 |

Table 2. Scaling behavior of indentation electromechanical constants

| Scaling parameter | $(\gamma c_{ij}, e_{ij}, \varepsilon_{ij})$, | | $(c_{ij}, \gamma e_{ij}, \varepsilon_{ij})$, | | $(c_{ij}, e_{ij}, \gamma \varepsilon_{ij})$, | |
|---|---|---|---|---|---|---|
| | $\gamma \to 0$ | $\gamma \to \infty$. | $\gamma \to 0$ | $\gamma \to \infty$. | $\gamma \to 0$ | $\gamma \to \infty$. |
| Indentation elastic stiffness, $C_1^*$ | $\gamma$ | $\gamma$ | *const* | *const* | *const* | *const* |
| Indentation piezocoelectric constant, $C_2^*$ | *const* | *const* | $\gamma$ | $\gamma$ | *const* | *const* |
| Indentation dielectric constant, $C_4^*$ | $\gamma^{-1}$ | *const* | *const* | $\gamma^2$ | *const* | $\gamma$ |
| Maximum potential, $\psi_{\max}$ | $\gamma$ | *const* | $\gamma$ | $\gamma^{-1}$ | *const* | $\gamma^{-1}$ |

Table 3. Asymptotic field behavior for mechanical problem (A)

| Function | $\rho = 0, z \to \infty$ | $z = 0, \rho \to \infty, \phi = 0$ |
|---|---|---|
| $u$ | 0 | $\frac{2H^*}{\pi R}\frac{2a^3}{3\rho}\sum_{j=1}^{3}\left(N_j^* C_1^* + L_j^* C_2^*\right)$ |
| $u_z$ | $-\frac{2H^*}{\pi R}\sum_{j=1}^{3}\frac{m_j^*}{\gamma_j^*}\left(N_j^* C_1^* + L_j^* C_2^*\right)\frac{2a^3}{3z_j}$ | $\frac{1}{\pi R}\frac{4a^3}{3\rho}$ |
| $\psi$ | $-\frac{2H^*}{\pi R}\sum_{j=1}^{3}\frac{k_j^*}{\gamma_j^*}\left(N_j^* C_1^* + L_j^* C_2^*\right)\frac{2a^3}{3z_j}$ | 0 |
| $\sigma_1$ | $\frac{8H^*}{\pi R}\sum_{j=1}^{3}\left(C_{66} - \frac{\alpha_j^*}{\gamma_j^{*2}}\right)\left(N_j^* C_1^* + L_j^* C_2^*\right)\frac{a^3}{3z_j^2}$ | 0 |
| $\sigma_2$ | 0 | $\frac{8C_{66}H^*}{3\pi R}\frac{2a^3}{\rho^2}\sum_{j=1}^{3}\left(N_j^* C_1^* + L_j^* C_2^*\right)$ |
| $\sigma_{zz}$ | $\frac{4H^*}{\pi R}\sum_{j=1}^{3}\alpha_j^*\left(N_j^* C_1^* + L_j^* C_2^*\right)\frac{a^3}{3z_j^2}$ | 0 |
| $\tau_z$ | 0 | 0 |
| $D_z$ | $\frac{4H^*}{\pi R}\sum_{j=1}^{3}\beta_j^*\left(N_j^* C_1^* + L_j^* C_2^*\right)\frac{a^3}{3z_j^2}$ | 0 |
| $D$ | 0 | $\frac{2H^*}{\pi R}\frac{2a^3}{3\rho^2}\sum_{j=1}^{3}\frac{\beta_j^*}{\gamma_j^*}\left(N_j^* C_1^* + L_j^* C_2^*\right)$ |

Table 4. Asymptotic behavior for electrical problem (B)

| Function | $\rho = 0, z \to \infty$ | $z = 0, \rho \to \infty, \phi = 0$ |
|---|---|---|
| $u$ | 0 | $-\frac{2\psi_0 H^*}{\pi}\frac{a}{\rho}\sum_{j=1}^{3}\left(N_j^* C_3^* + L_j^* C_4^*\right)$ |
| $u_z$ | $-\frac{2\psi_0 H^*}{\pi}\sum_{j=1}^{3}\frac{m_j^*}{\gamma_j^*}\left(N_j^* C_3^* + L_j^* C_4^*\right)\frac{a}{z_j}$ | 0 |
| $\psi$ | $-\frac{2\psi_0 H^*}{\pi}\sum_{j=1}^{3}\frac{k_j^*}{\gamma_j^*}\left(N_j^* C_3^* + L_j^* C_4^*\right)\frac{a}{z_j}$ | $\frac{2\psi_0}{\pi}\frac{a}{\rho}$ |
| $\sigma_1$ | $\frac{4\psi_0 H^*}{\pi}\sum_{j=1}^{3}\left(C_{66} - \frac{\alpha_j^*}{\gamma_j^{*2}}\right)\left(N_j^* C_3^* + L_j^* C_4^*\right)\frac{a}{z_j^2}$ | 0 |
| $\sigma_2$ | 0 | $\frac{8H^* C_{66}\psi_0}{\pi}\frac{a}{\rho^2}\sum_{j=1}^{3}\left(N_j^* C_3^* + L_j^* C_4^*\right)$ |
| $\sigma_{zz}$ | $\frac{2\psi_0 H^*}{\pi}\sum_{j=1}^{3}\alpha_j^*\left(N_j^* C_3^* + L_j^* C_4^*\right)\frac{a}{z_j^2}$ | 0 |
| $\tau_z$ | 0 | 0 |
| $D_z$ | $\frac{2\psi_0 H^*}{\pi}\sum_{j=1}^{3}\beta_j^*\left(N_j^* C_3^* + L_j^* C_4^*\right)\frac{a}{z_j^2}$ | 0 |
| $D$ | 0 | $\frac{2\psi_0 H^*}{\pi}\frac{a}{\rho^2}\sum_{j=1}^{3}\frac{\beta_j^*}{\gamma_j^*}\left(N_j^* C_3^* + L_j^* C_4^*\right)$ |

## Appendix A

The following notations and complex combinations are used for displacements, stresses, the electric potential, and the electric displacement components:

$$u \equiv u_x + iu_y , \quad u_z , \quad \psi , \quad D \equiv D_x + iD_y , \quad D_z \tag{A.1}$$

$$\sigma_1 \equiv \sigma_{xx} + \sigma_{yy} , \quad \sigma_2 \equiv \sigma_{xx} - \sigma_{yy} + 2i\sigma_{xy} , \quad \sigma_{zz} , \quad \tau_z \equiv \sigma_{zx} + i\sigma_{yz} \tag{A.2}$$

where $c_{ij}$ denote transversely isotropic elastic stiffnesses, $e_{ij}$ – piezoelectric constants, $\varepsilon_{ij}$ – dielectric permeabilities and $\alpha_j^* = c_{44}\left(1+m_j^*\right)+e_{15}k_j^*$, $\beta_j^* = e_{15}\left(1+m_j^*\right)-\varepsilon_{11}k_j^*$ $\left(j=1,2,3\right)$. Constants $m_j^*$, $k_j^*$ $\left(j=1,2,3\right)$ in the text to follow are obtained to be as follows:

$$m_j = \frac{\left(c_{11}\gamma_j^2 - c_{44}\right)\left(\varepsilon_{33} - \gamma_j^2\varepsilon_{11}\right) + \gamma_j^2\left(e_{15}+e_{31}\right)^2}{\left(e_{33} - \gamma_j^2 e_{15}\right)\left(e_{15}+e_{31}\right) + \left(c_{13}+c_{44}\right)\left(\varepsilon_{33} - \gamma_j^2\varepsilon_{11}\right)} \tag{A.3}$$

$$k_j = \frac{\left(c_{11}\gamma_j^2 - c_{44}\right)\left(e_{33} - \gamma_j^2 e_{15}\right) - \gamma_j^2\left(c_{13}+c_{44}\right)\left(e_{15}+e_{31}\right)}{\left(e_{33} - \gamma_j^2 e_{15}\right)\left(e_{15}+e_{31}\right) + \left(c_{13}+c_{44}\right)\left(\varepsilon_{33} - \gamma_j^2\varepsilon_{11}\right)} \tag{A.4}$$

where $\gamma_j^{*2} = \lambda_j$ are roots of the cubic equation

$$A\lambda_j^3 - B\lambda_j^2 + C\lambda_j - D = 0 \tag{A.5}$$

with coefficients

$$A = c_{11}\left(c_{44}\varepsilon_{11} + e_{15}^2\right) \tag{A.6}$$

$$\begin{aligned} B = {} & c_{44}\left[c_{11}\varepsilon_{33} + \left(e_{15}+e_{31}\right)^2\right] + \varepsilon_{11}\left[c_{11}c_{33} + c_{44}^2 - \left(c_{13}+c_{44}\right)^2\right] \\ & + 2e_{15}\left[c_{11}e_{33} - \left(c_{13}+c_{44}\right)\left(e_{15}+e_{31}\right)\right] + c_{44}e_{15}^2 \end{aligned} \tag{A.7}$$

$$\begin{aligned} C = {} & c_{33}\left[c_{44}\varepsilon_{11} + \left(e_{15}+e_{31}\right)^2\right] + \varepsilon_{33}\left[c_{11}c_{33} + c_{44}^2 - \left(c_{13}+c_{44}\right)^2\right] \\ & + 2e_{33}\left[c_{44}e_{15} - \left(c_{13}+c_{44}\right)\left(e_{15}+e_{31}\right)\right] + c_{11}e_{33}^2 \end{aligned} \tag{A.8}$$

$$D = c_{44}\left(c_{33}\varepsilon_{33} + e_{33}^2\right) \tag{A.9}$$

Constants $m_j^*$ and $k_j^*$ can be expressed in terms of roots $\lambda_j$ (formula (2.6) in Ref.[64]) Of the six roots for $\gamma$ that correspond to three roots for $\lambda$, obtained from Eq. (A5), the roots $\gamma_{1,2,3}$ that have positive real parts must be chosen, to ensure that displacements are real.

The following combinations of the piezoelectric constants are used:

$$H^* = \frac{1}{2\pi\left(e_{15}^2 + c_{44}\varepsilon_{11}\right)\sum_{j=1}^{3}\frac{\alpha_j^* a_j^*}{\gamma_j^{*2}}} = -\frac{1}{2\pi\sum_{j=1}^{3}\alpha_j^* N_j^*} \tag{A.10}$$

$$N_1^* = \frac{\alpha_3^* \beta_2^*}{\gamma_3^*} - \frac{\alpha_2^* \beta_3^*}{\gamma_2^*}, \tag{A.11}$$

$$L_1^* = \frac{\alpha_3^* \alpha_2^*}{\gamma_3^*} - \frac{\alpha_2^* \alpha_3^*}{\gamma_2^*}, \tag{A.12}$$

$$a_1^* = \gamma_1^*\left[\left(1 + m_2^*\right)k_3^* - \left(1 + m_3^*\right)k_2^*\right] \ . \tag{A.13}$$

Values for other constants $N_j^*$, $L_j^*$, $a_j^*$ are obtained by cyclic permutation of indices as $1 \to 2 \to 3 \to 1$. The following geometric parameters ($j = 1,2,3$) are used:

$$2l_{1j}(z) = \sqrt{(a+\rho)^2 + z_j^2} - \sqrt{(a-\rho)^2 + z_j^2}, \tag{A.14}$$

$$2l_{2j}(z) = \sqrt{(a+\rho)^2 + z_j^2} + \sqrt{(a-\rho)^2 + z_j^2}, \tag{A.15}$$

$$z_j = z/\gamma_j \tag{A.16}$$

The constants in Eqs. (17-20) are defined as

$$C_1^* = -\frac{1}{B^*}\sum_{j=1}^{3}\frac{k_j^*}{\gamma_j^*}L_j^*, \tag{A.17}$$

$$C_2^* = \frac{1}{B^*} \sum_{j=1}^{3} \frac{k_j^*}{\gamma_j^*} N_j^* \, , \qquad \text{(A. 18)}$$

$$C_3^* = \frac{1}{B^*} \sum_{j=1}^{3} \frac{m_j^*}{\gamma_j^*} L_j^* \, , \qquad \text{(A.19)}$$

$$C_4^* = -\frac{1}{B^*} \sum_{j=1}^{3} \frac{m_j^*}{\gamma_j^*} N_j^* \qquad \text{(A.20)}$$

$$B^* = H^* \left[ \sum_{j=1}^{3} \frac{m_j^*}{\gamma_j^*} N_j^* \sum_{i=1}^{3} \frac{k_i^*}{\gamma_i^*} L_i^* - \sum_{j=1}^{3} \frac{m_j^*}{\gamma_j^*} L_j^* \sum_{i=1}^{3} \frac{k_i^*}{\gamma_i^*} N_i^* \right] \qquad \text{(A.21)}$$

*Remark.* Parameters $H^*$, $N_i^*$, $L_i^*$, that will enter the solution, are explicitly expressed in terms of the piezoelectric constants – in contrast with the solution of Chen (Ref.[51]) where the dependence of the solution on the piezoelectric constants is not explicit.

## Appendix B

The solution of the sub-problem (A) is as follows:

$$u=-\frac{2H^*}{\pi R}\rho e^{i\phi}\sum_{j=1}^{3}\left(N_j^*C_1^*+L_j^*C_2^*\right)\left[-z_j\arcsin\left(\frac{l_{1j}}{\rho}\right)+\left(a^2-l_{1j}^2\right)^{1/2}\left(1-\frac{l_{1j}^2+2a^2}{3\rho^2}\right)+\frac{2a^3}{3\rho^2}\right] \quad \text{(B.1)}$$

$$\sigma_1=-\frac{8H^*}{\pi R}\sum_{j=1}^{3}\left(c_{66}-\frac{\alpha_j^*}{\gamma_j^{*2}}\right)\left(N_j^*C_1^*+L_j^*C_2^*\right)\left[z_j\arcsin\left(\frac{l_{1j}}{\rho}\right)-\left(a^2-l_{1j}^2\right)^{1/2}\right] \quad \text{(B.2)}$$

$$\sigma_2=\frac{8C_{66}H^*}{3\pi R}\frac{e^{2i\phi}}{\rho^2}\sum_{j=1}^{3}\left(N_j^*C_1^*+L_j^*C_2^*\right)\left[2a^3-\left(l_{1j}^2+2a^2\right)\left(a^2-l_{1j}^2\right)^{1/2}\right] \quad \text{(B.3)}$$

$$\sigma_{zz}=-\frac{4H^*}{\pi R}\sum_{j=1}^{3}\alpha_j^*\left(N_j^*C_1^*+L_j^*C_2^*\right)\left[z_j\arcsin\left(\frac{l_{1j}}{\rho}\right)-\left(a^2-l_{1j}^2\right)^{1/2}\right] \quad \text{(B.4)}$$

$$\tau_z=-\frac{2H^*}{\pi R}\rho e^{i\phi}\sum_{j=1}^{3}\frac{\alpha_j^*}{\gamma_j^*}\left(N_j^*C_1^*+L_j^*C_2^*\right)\left[-\arcsin\left(\frac{l_{1j}}{\rho}\right)+\frac{a\left(l_{2j}^2-a^2\right)^{1/2}}{l_{2j}^2}\right] \quad \text{(B.5)}$$

$$D_z=-\frac{4H^*}{\pi R}\sum_{j=1}^{3}\beta_j^*\left(N_j^*C_1^*+L_j^*C_2^*\right)\left[z_j\arcsin\left(\frac{l_{1j}}{\rho}\right)-\left(a^2-l_{1j}^2\right)^{1/2}\right] \quad \text{(B.6)}$$

$$D=-\frac{2H^*}{\pi R}\rho e^{i\phi}\sum_{j=1}^{3}\frac{\beta_j^*}{\gamma_j^*}\left(N_j^*C_1^*+L_j^*C_2^*\right)\left[-\arcsin\left(\frac{l_{1j}}{\rho}\right)+\frac{a\left(l_{2j}^2-a^2\right)^{1/2}}{l_{2j}^2}\right] \quad \text{(B.7)}$$

The solution of the sub-problem (B) is as follows:

$$u=-\frac{2\psi_0H^*}{\pi}\sum_{j=1}^{3}\left(N_j^*C_3^*+L_j^*C_4^*\right)\frac{ae^{i\phi}}{\rho}\left[1-\frac{\left(a^2-l_{1j}^2\right)^{1/2}}{a}\right] \quad \text{(B.8)}$$

$$\sigma_1=\frac{4\psi_0H^*}{\pi}\sum_{j=1}^{3}\left(c_{66}-\frac{\alpha_j^*}{\gamma_j^{*2}}\right)\left(N_j^*C_3^*+L_j^*C_4^*\right)\frac{\left(a^2-l_{1j}^2\right)^{1/2}}{l_{2j}^2-l_{1j}^2} \quad \text{(B.9)}$$

$$\sigma_2 = -\frac{4\psi_0 H^* C_{66} e^{2i\phi}}{\pi} \sum_{j=1}^{3} \left(N_j^* C_3^* + L_j^* C_4^*\right) \left\{ \frac{\left(a^2 - l_{1j}^2\right)^{1/2}}{l_{2j}^2 - l_{1j}^2} - \frac{2a}{\rho^2} \left[ 1 - \frac{\left(a^2 - l_{1j}^2\right)^{1/2}}{a} \right] \right\} \quad \text{(B.10)}$$

$$\sigma_{zz} = \frac{2\psi_0 H^*}{\pi} \sum_{j=1}^{3} \alpha_j^* \left(N_j^* C_3^* + L_j^* C_4^*\right) \frac{\left(a^2 - l_{1j}^2\right)^{1/2}}{l_{2j}^2 - l_{1j}^2} \quad \text{(B.11)}$$

$$\tau_z = \frac{2\psi_0 H^* e^{i\phi}}{\pi} \sum_{j=1}^{3} \frac{\alpha_j^*}{\gamma_j^*} \left(N_j^* C_3^* + L_j^* C_4^*\right) \frac{l_{1j}\left(l_{2j}^2 - a^2\right)^{1/2}}{l_{2j}\left(l_{2j}^2 - l_{1j}^2\right)} \quad \text{(B.12)}$$

$$D_z = \frac{2\psi_0 H^*}{\pi} \sum_{j=1}^{3} \beta_j^* \left(N_j^* C_3^* + L_j^* C_4^*\right) \frac{\left(a^2 - l_{1j}^2\right)^{1/2}}{l_{2j}^2 - l_{1j}^2} \quad \text{(B.13)}$$

$$D = \frac{2\psi_0 H^* e^{i\phi}}{\pi} \sum_{j=1}^{3} \frac{\beta_j^*}{\gamma_j^*} \left(N_j^* C_3^* + L_j^* C_4^*\right) \frac{l_{1j}\left(l_{2j}^2 - a^2\right)^{1/2}}{l_{2j}\left(l_{2j}^2 - l_{1j}^2\right)} \quad \text{(B.14)}$$

## Figure Captions

**Figure 1.** Schematic representation of the PFM experiment in the weak (a) and strong (b) indentation regimes. In the weak indentation case, the indentation force and contact area is small, and the field inside the material can be determined from the electrostatic sphere-plane model. In the strong indentation regime, the capacitance of the contact area dominates over the capacitance of the spherical part of the tip and determines the field inside the material. In this regime, the elastic and electroelastic effects due to indentation force are significant and should be taken into account. (c) Equivalent circuit for tip-surface junction. Shown are contributions from spherical part of the tip and contact area with the quantum capacitance limit taken into account. (d) Schematic illustration of the contact, sphere and non-local contributions to the field.

**Figure 2 (Color online).** (a) Dielectric constant dependence of critical ratio of tip radius to contact radius for which capacitances of spherical part of the tip and contact area are equal. (b) Ratio of tip radius to contact radius in Hertzian model for different indentation forces.

**Figure 3 (Color online).** (a) Dielectric constant dependence of tip-surface capacitance and effective charge-surface separation. (b) Potential distribution in the ferroelectric in the sphere-plane and point charge models for different charge-surface separations. Note the differences in the field distributions in the vicinity of the contact despite identical asymptotic behavior.

**Figure 4 (Color online).** Potential distribution inside the material calculated in the weak indentation regime using sphere plane model for tip radius $R$ = 50 nm and in the strong indentation regime for several contact diameters calculated using exact model in Section V for PZT6b. Also shown is the electroelastic contribution to potential due to the indentation force.

**Figure 5.** Sensitivity function of the coupling coefficients in stiffness relations Eqs. (23,24) in the ($c_{ij}$, $e_{ij}$, $\varepsilon_{ij}$) and ($s_{ij}$, $d_{ij}$, $\varepsilon_{ij}$) representations calculated for PZT6b.

**Figure 6.** Scaling behavior of indentation elastic stiffness (a), indentation piezocoefficient (b), indentation dielectric constant (c) and maximum induced potential (d) for piezoelectric scaling.

**Figure 7.** Scaling behavior of indentation elastic stiffness (a), indentation piezocoefficient (b), indentation dielectric constant (c) and maximum induced potential (d) for elastic and dielectric scaling.

**Figure 8.** (a) Relative contributions of elastic, electroelastic and electrostatic components to the total force and charge in stiffness relations Eqs. (23,24) for $BaTiO_3$ for spherical indenter with $R$ = 50 nm. In region I, the indentor charge is dominated by the electroelastic contribution and the force is determined by the elastic contribution. In region II, indentor charge is dominated by the electrostatic contribution and the force is determined by the elastic contribution. In region III, indentor charge is dominated by the electrostatic contribution and the force is determined by the electroelastic contribution. (b) Response diagram as function of indentation force and tip bias.

**Figure 9.** (a) Bias dependence of indentation depth, (b) contact radius, (c) piezoresponse amplitude, and (d) effective dielectric constant for the PZT6b and tip radius $R$ = 50 nm for indentation force 10 nN (solid), 100 nN (dash), 1 μN (dash dot), and 10 μN (short dash).

**Figure 10.** Sensitivity function of the piezoresponse amplitude $A_{piezo}$ in the ($c_{ij}$, $e_{ij}$, $\varepsilon_{ij}$) representation (a) and ($s_{ij}$, $d_{ij}$, $\varepsilon_{ij}$) representations (b) calculated for PZT6b. Normal component of electric field is related to the vertical strain component by piezoelectric constant $d_{33}$ (c). In the spherical indentation geometry, an additional contribution to response amplitude originates from the lateral component of electric fields related to the vertical strain component by piezoelectric constant $d_{15}$ (d). The ratio between the lateral and vertical field components is determined by $\varepsilon_{11}/\varepsilon_{33}$, thus rationalizing the dominant contributions of these constants to the sensitivity function for $A_{piezo}$.

**Figure 11 (Color online).** (a) Correlation between $A_{piezo}$ and $d_{33}$ and (b) correlation between effective dielectric constant and $\sqrt{\varepsilon_{11}\varepsilon_{33}}$ for several polycrystalline (■) and single crystal (▲) ferroelectric materials.

**Figure 12.** 2D spatial distribution of the electrostatic potential (a,b), normal displacement (c,d), normal stress (e,f) and electric displacement (g,h) in the electrical (a,c,e,g) and mechanical (b,d,f,h) sub-problems for contact radius $a$ = 3 nm, tip radius of curvature $R$ = 50

nm and tip potential $\Psi_0 = 1$ V for $BaTiO_3$. These conditions correspond to indentation force $P = 92.44$ nN.

**Figure 13.** 2D spatial distribution of the electrostatic potential (a,b), normal displacement (c,d), normal stress (e,f) and electric displacement (g,h) in the electrical (a,c,e,g) and mechanical (b,d,f,h) sub-problems for contact radius $a = 3$ nm, tip radius of curvature $R = 50$ nm and tip potential $\Psi_0 = 1$ V for $LiNbO_3$. These conditions correspond to indentation force $P = 148.3$ nN.

**Figure 14 (Color online).** Normal distance dependence of potential (a,d) and normal displacement (b,c) for the electrical (a,c) and mechanical (b,d) problems calculated for $R = 50$ nm, $a = 3$ nm and $\Psi_0 = 1$ V for $LiNbO_3$ (solid line), $BaTiO_3$ (dash), $LiTaO_3$ (dash dot), and PZT6b (dotted line).

**Figure 15 (Color online).** Normal distance dependence of normal stress $\sigma_{zz}$ (a,b) and normal component of the displacement vector $D_z$ (c,d) for the electrical (a,c) and mechanical (b,d) problems calculated for $R = 50$ nm, $a = 3$ nm, and $\Psi_0 = 1$ V for $LiNbO_3$ (solid line), $BaTiO_3$ (dash), $LiTaO_3$ (dash dot) and PZT6b (dotted line).

**Figure 16 (Color online).** Radial dependence of (a) potential, (b) lateral displacement, (c) normal stress and (d) electric displacement for the electrical problem calculated for $R = 50$ nm, $a = 3$ nm and $\Psi_0 = 1$ V for $LiNbO_3$ (solid line), $BaTiO_3$ (dash), $LiTaO_3$ (dash dot), and PZT6b (dotted line).

**Figure 17 (Color online).** Radial dependence of (a) normal displacement, (b) lateral displacement, (c) normal stress and (d) electric displacement for the mechanical problem calculated for $R$ = 50 nm, $a$ = 3 nm and $\Psi_0$ = 1 V for $LiNbO_3$ (solid line), $BaTiO_3$ (dash), $LiTaO_3$ (dash dot), and PZT6b (dotted line).

**Figure 18.** (a) The ratio of the potential distributions calculated for the point charge model and rigorous solution for $BaTiO_3$. (b) The ratio of the normal displacement distributions calculated for the point force model and rigorous solution for $BaTiO_3$. Note that the difference between rigorous and point charge/force solutions doesn't exceed 50% for distances as small as ~ 0.3 $a$, justifying use of the point charge approximation for certain bias-induced phenomena in ferroelectric materials.

**Figure 19.** Signal generation volume in PFM for (a) $BaTiO_3$ and (b) $LiNbO_3$ for $R$ = 50 nm, $a$ = 3 nm and $\Psi_0$ = 0 V.

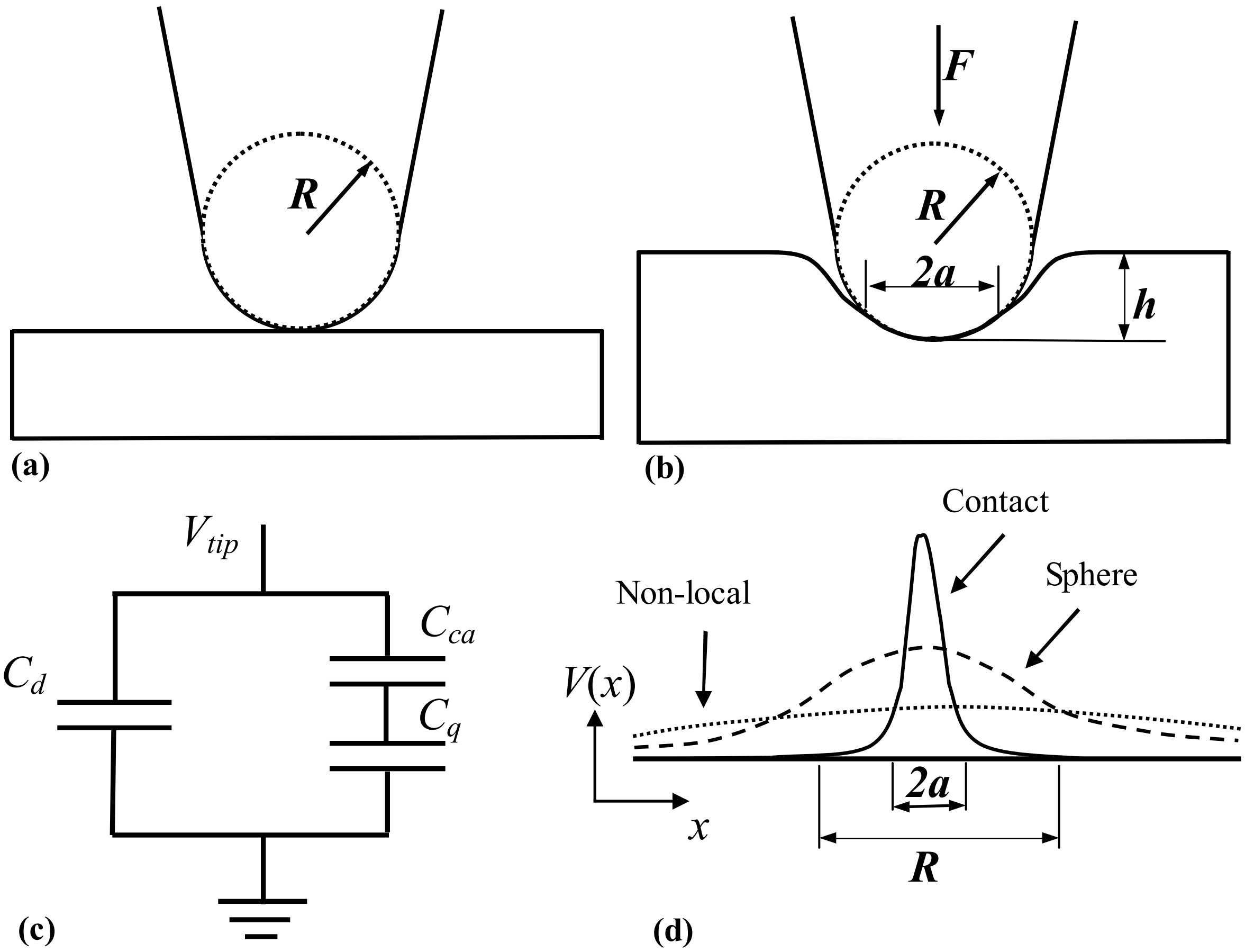

**Fig. 1.** S.V. Kalinin, E. Karapetian, and M. Kachanov

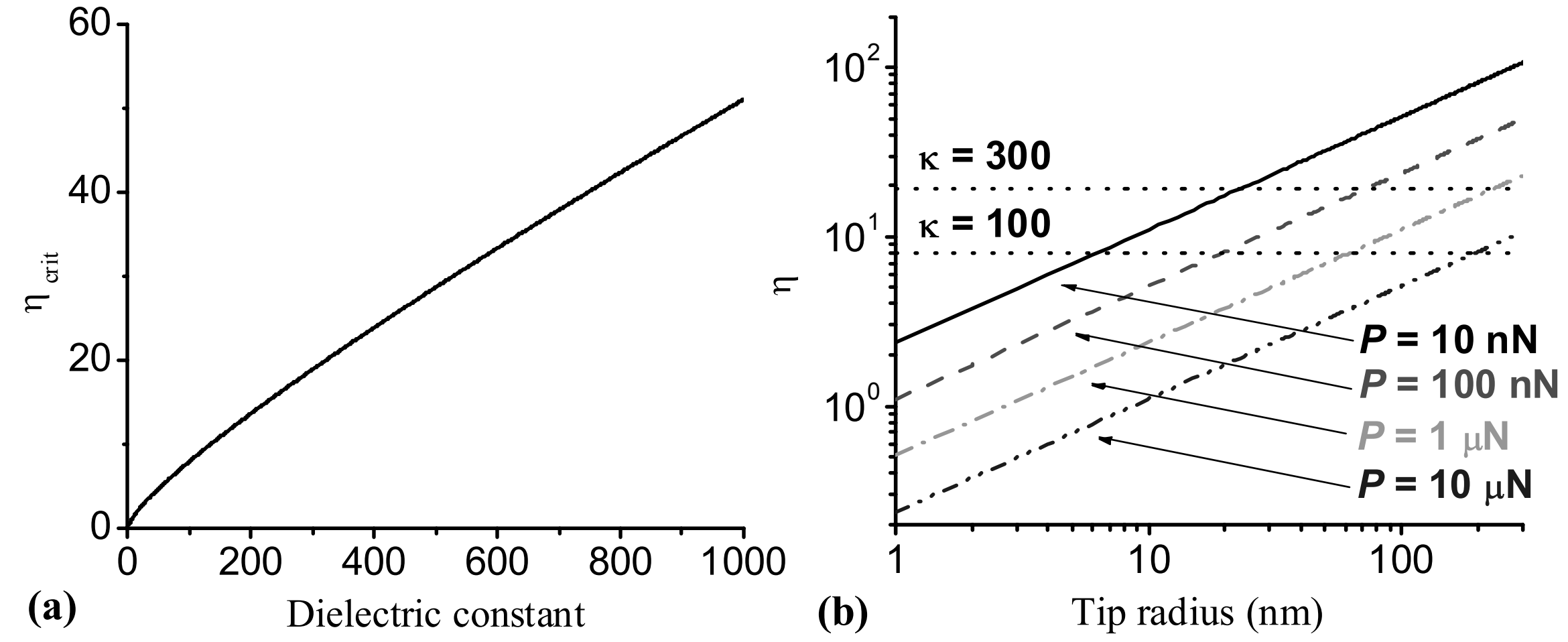

**Fig. 2.** S.V. Kalinin, E. Karapetian, and M. Kachanov

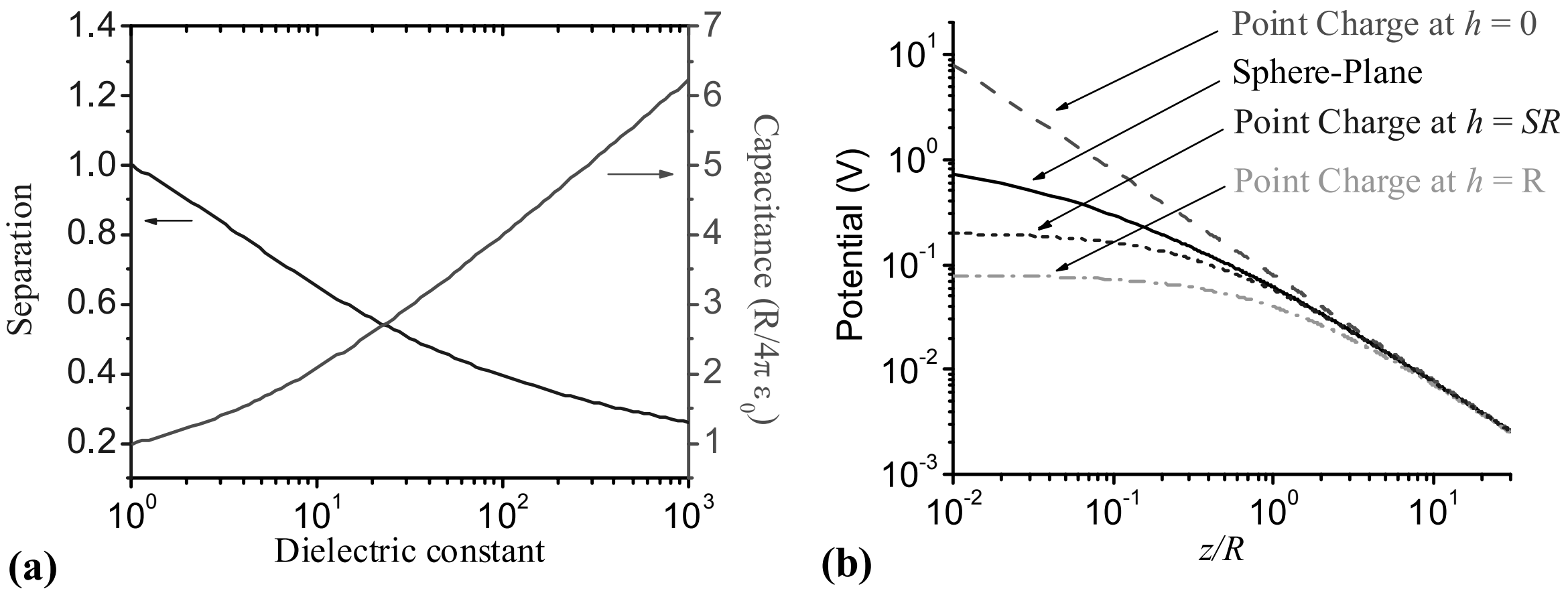

**Fig. 3.** S.V. Kalinin, E. Karapetian, and M. Kachanov

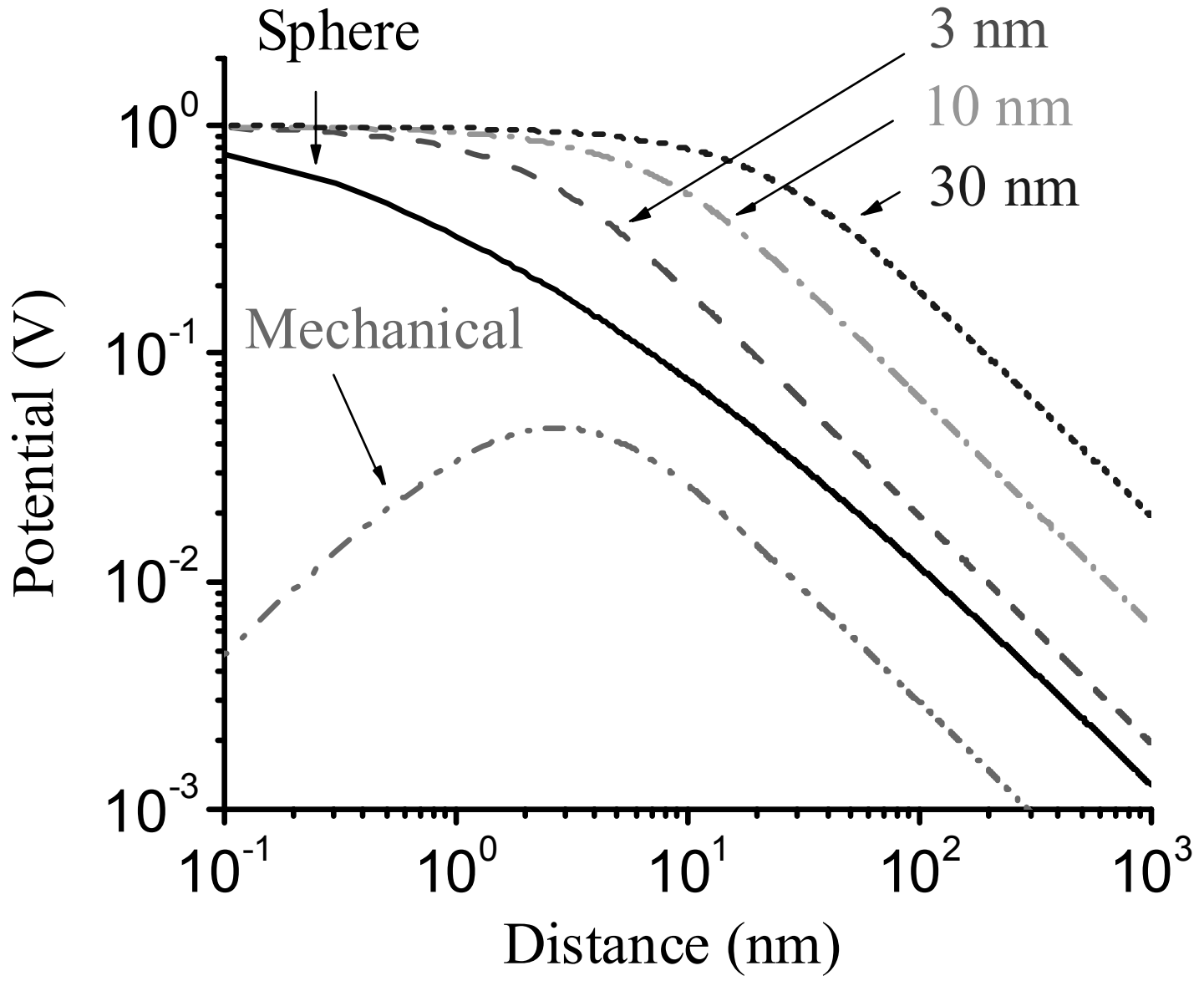

**Fig. 4.** S.V. Kalinin, E. Karapetian, and M. Kachanov

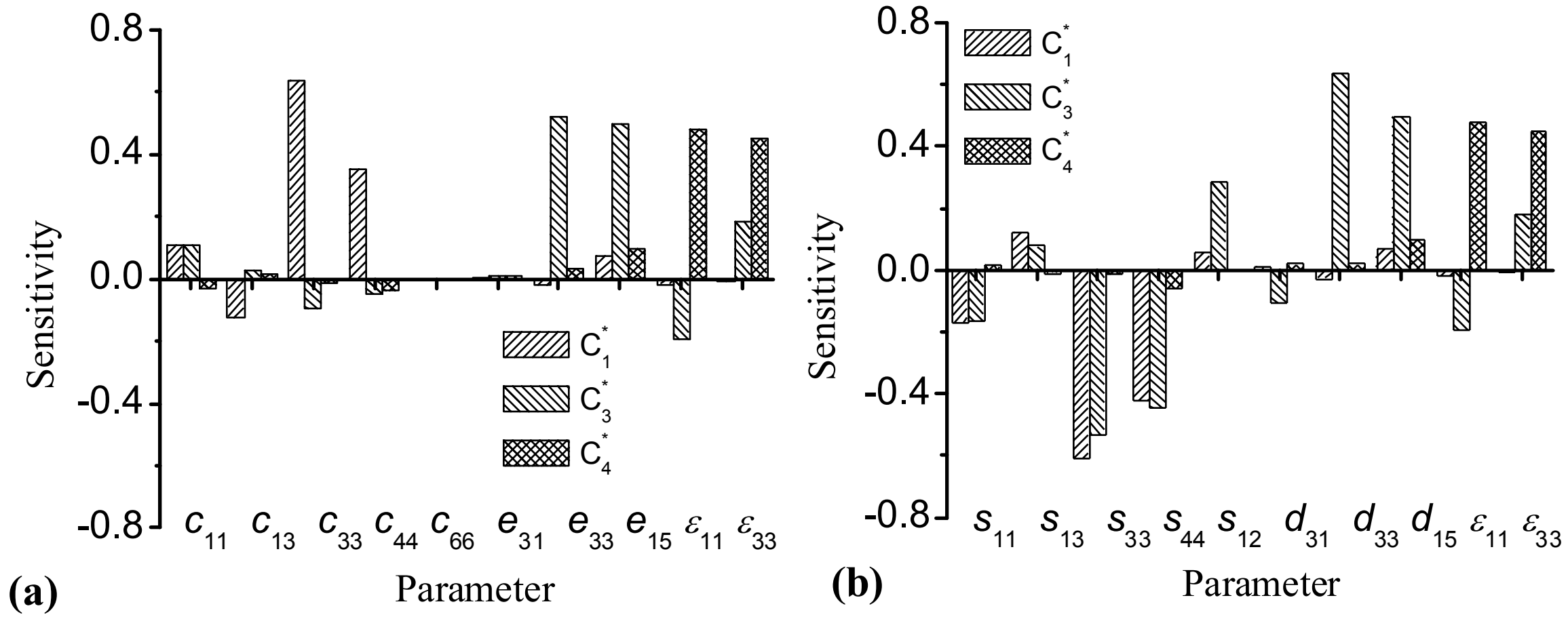

**Fig. 5.** S.V. Kalinin, E. Karapetian, and M. Kachanov

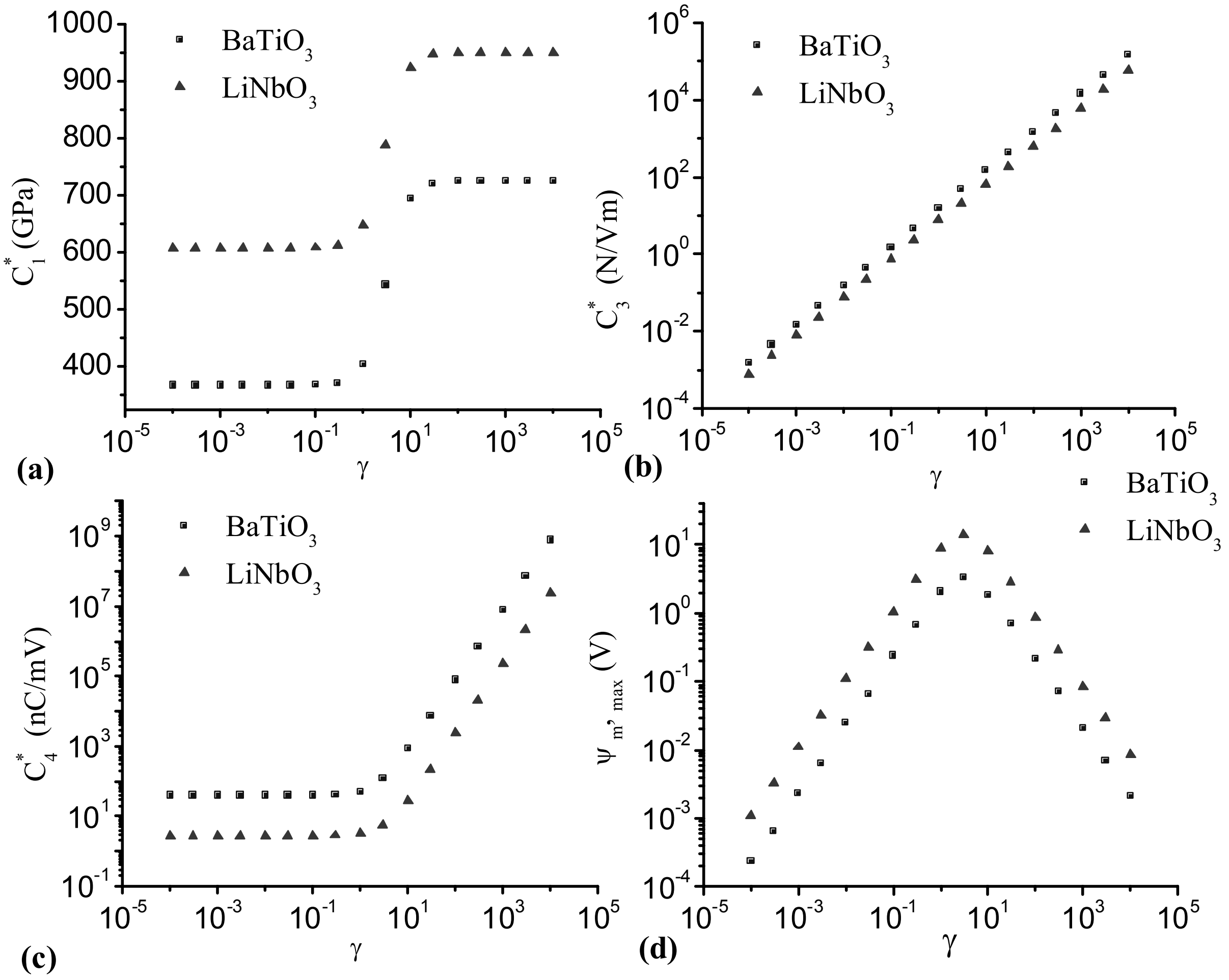

**Fig. 6.** S.V. Kalinin, E. Karapetian, and M. Kachanov

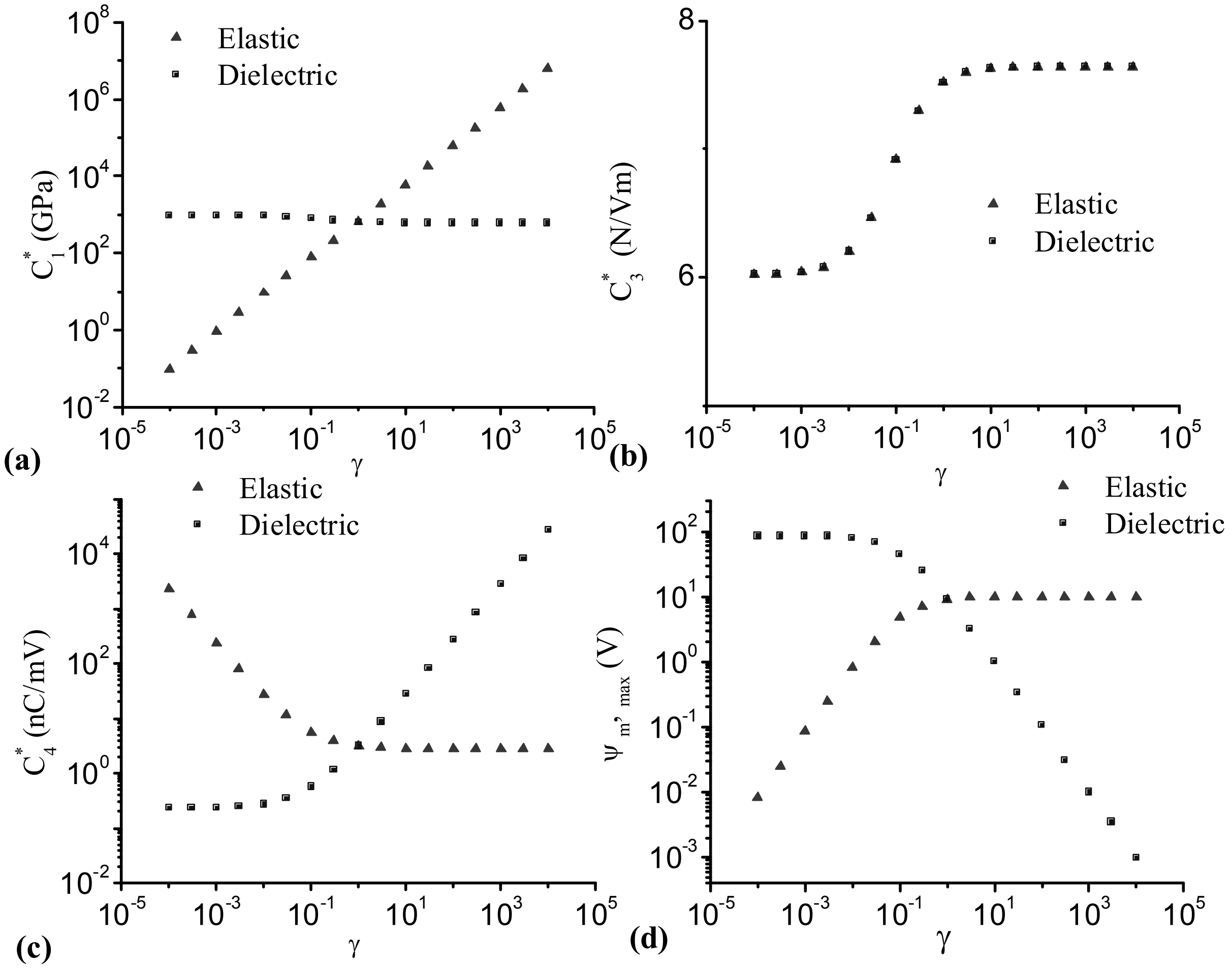

**Fig. 7.** S.V. Kalinin, E. Karapetian, and M. Kachanov

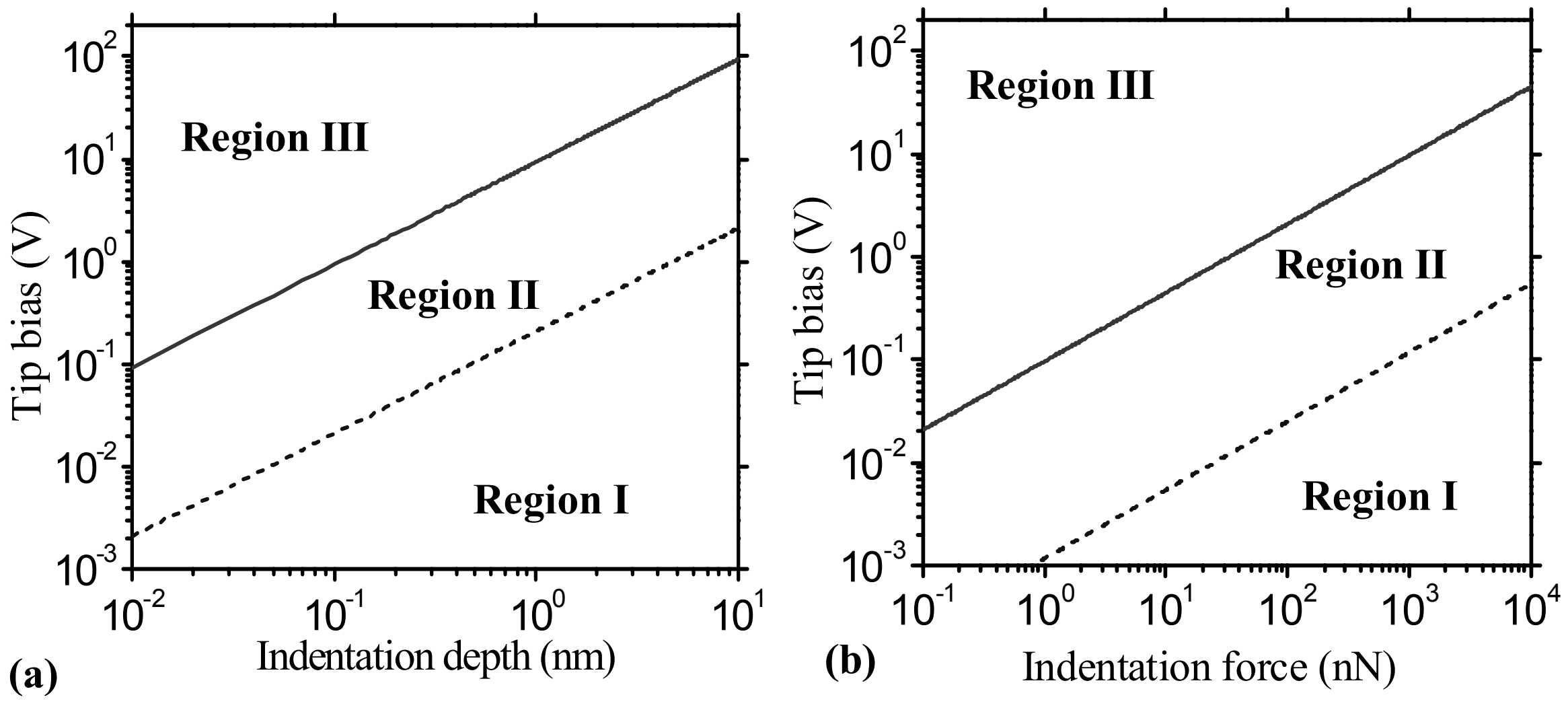

**Fig. 8.** S.V. Kalinin, E. Karapetian, and M. Kachanov

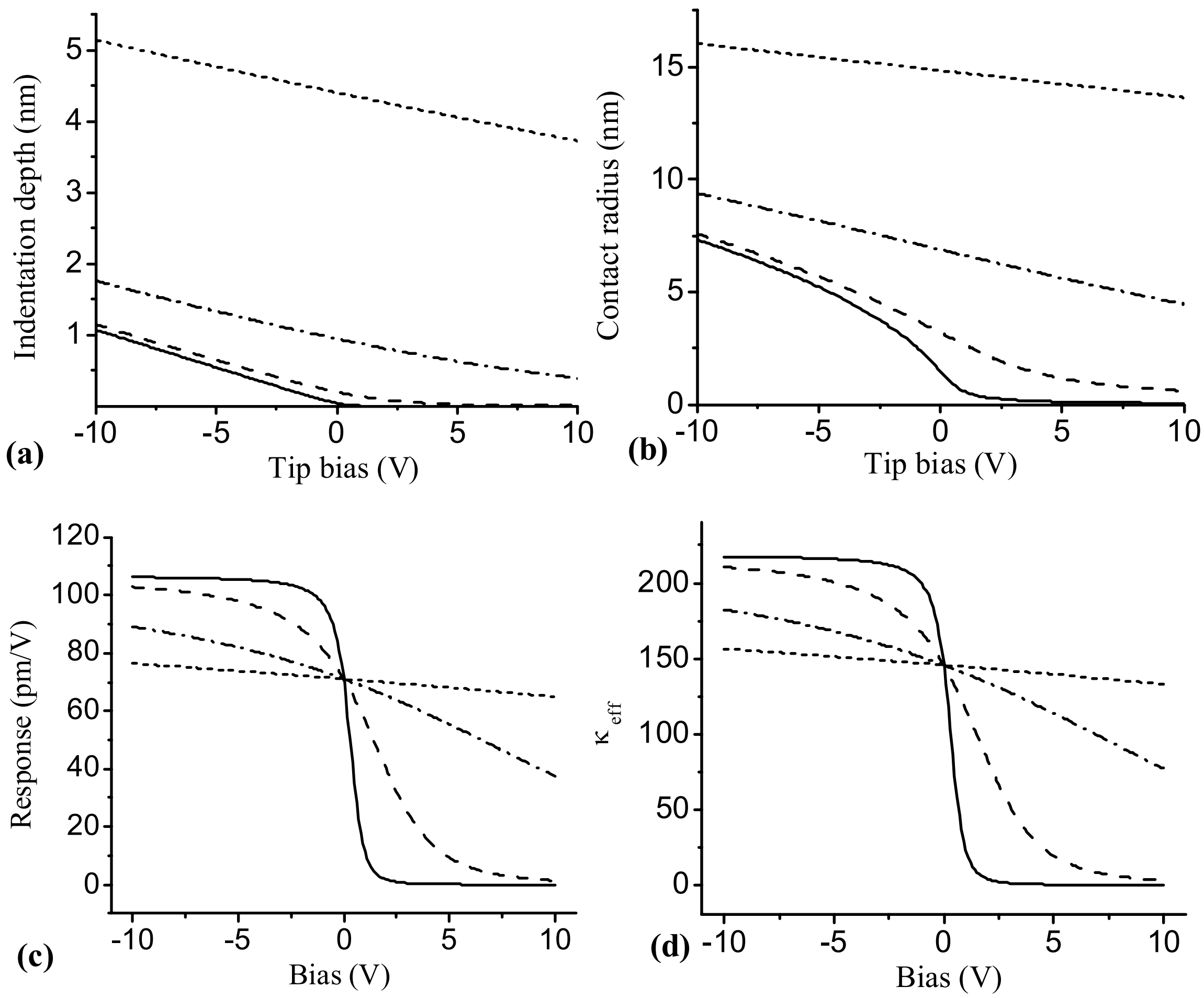

**Fig. 9.** S.V. Kalinin, E. Karapetian, and M. Kachanov

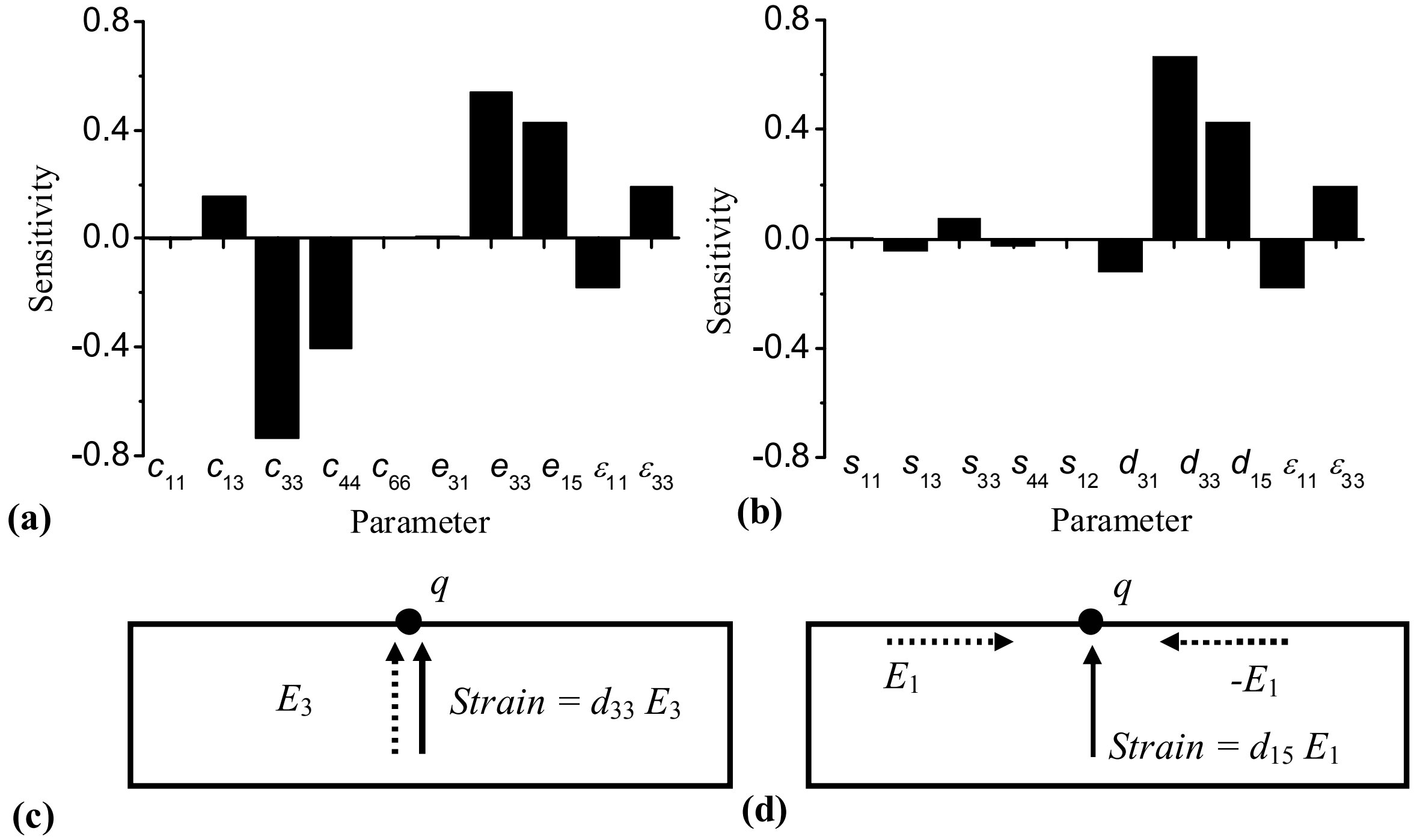

**Fig. 10.** S.V. Kalinin, E. Karapetian, and M. Kachanov

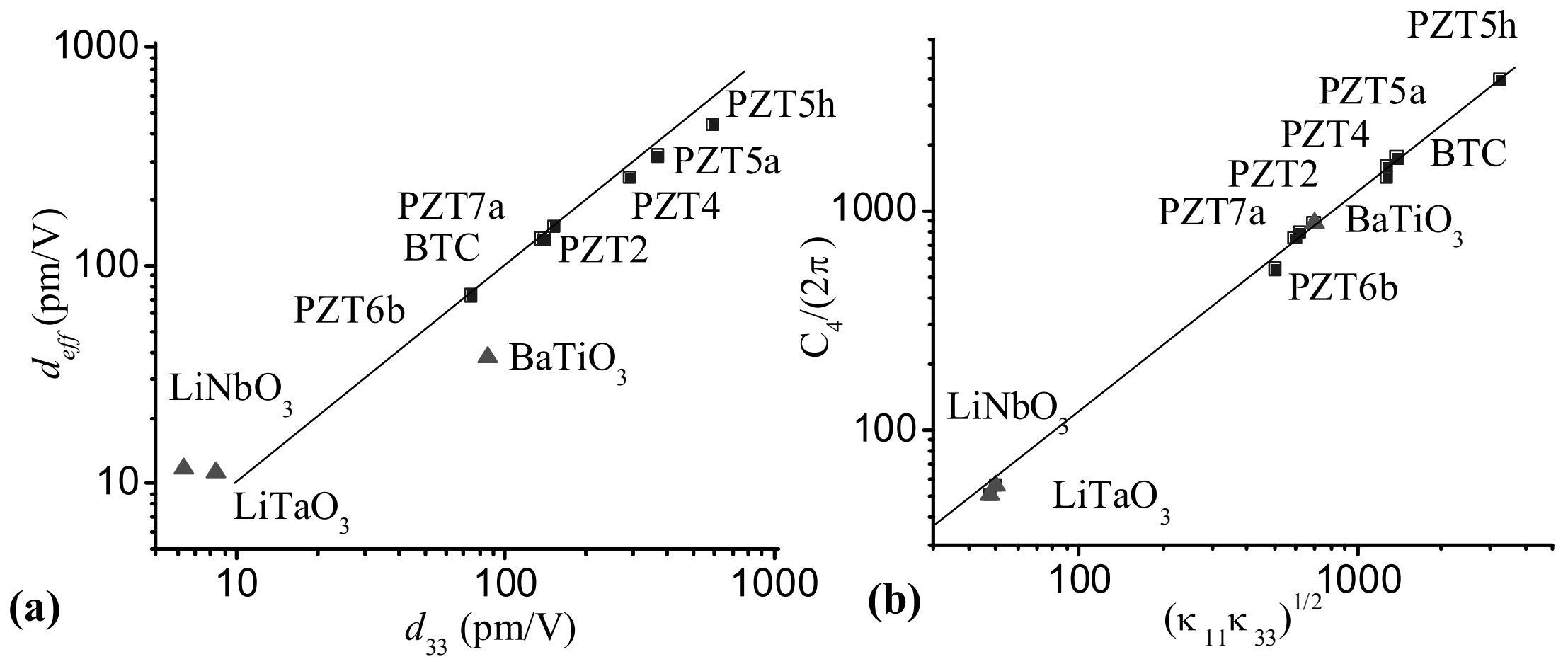

**Fig. 11.** S.V. Kalinin, E. Karapetian, and M. Kachanov

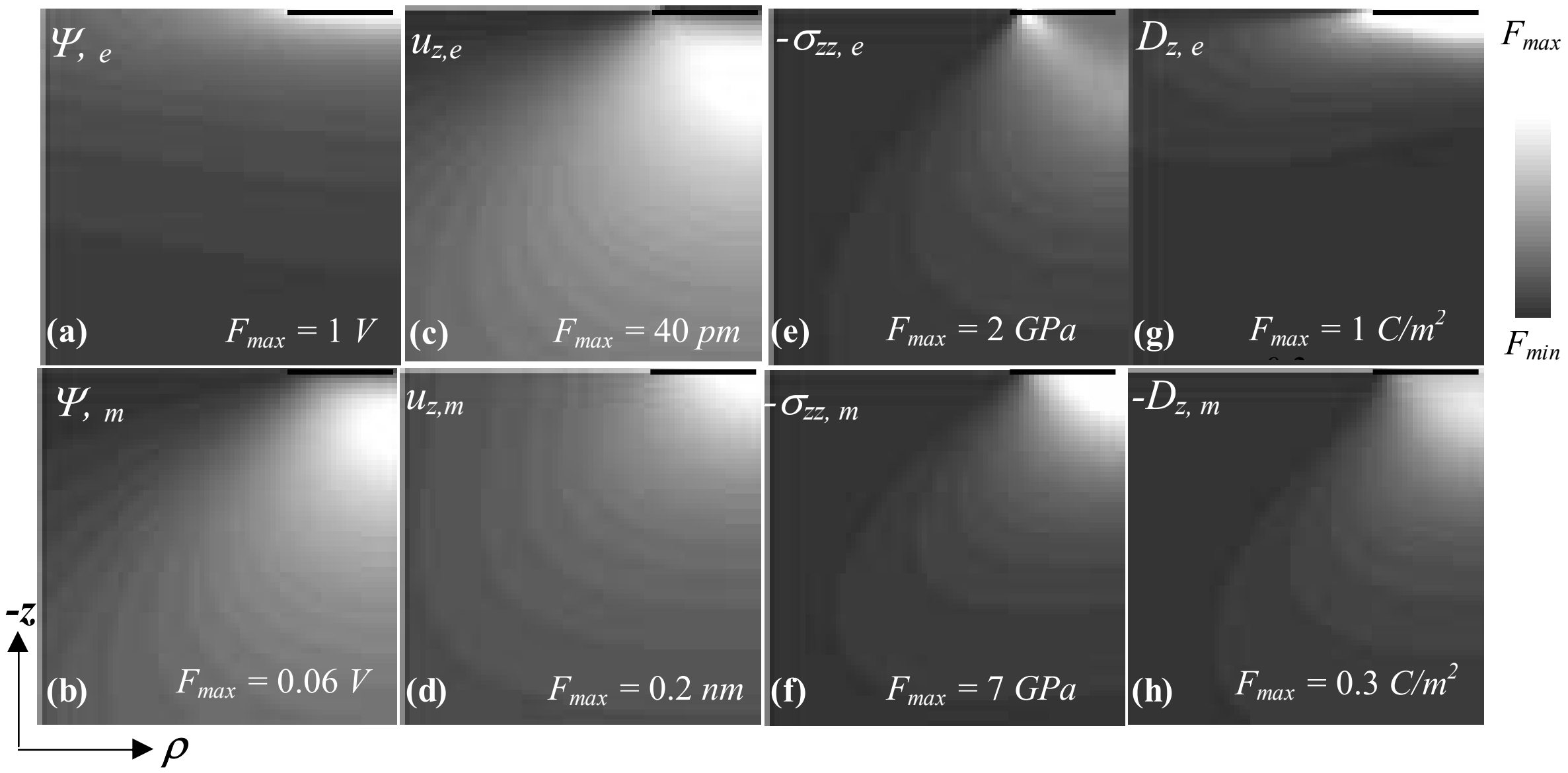

**Fig. 12.** S.V. Kalinin, E. Karapetian, and M. Kachanov

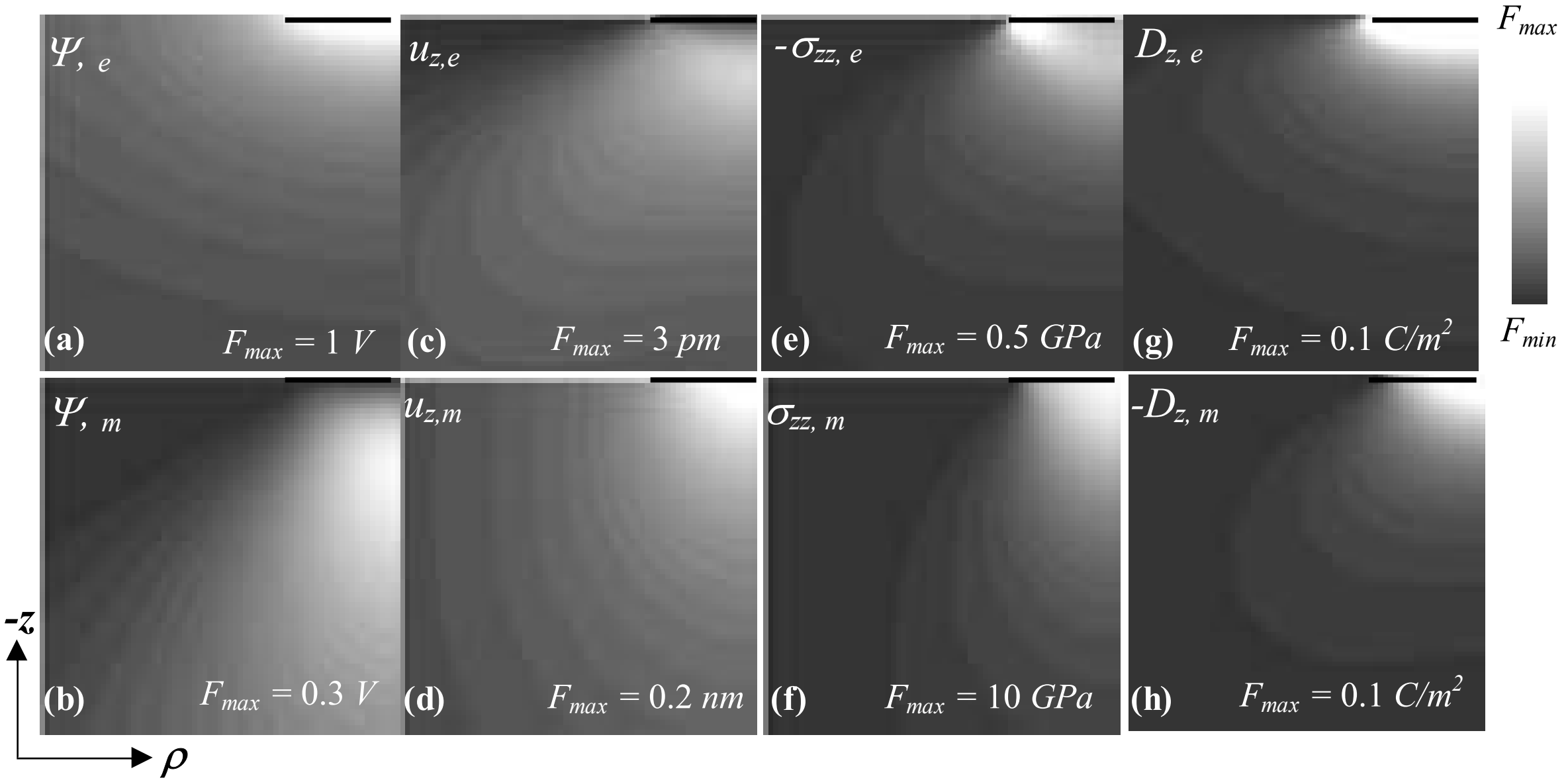

**Fig. 13.** S.V. Kalinin, E. Karapetian, and M. Kachanov

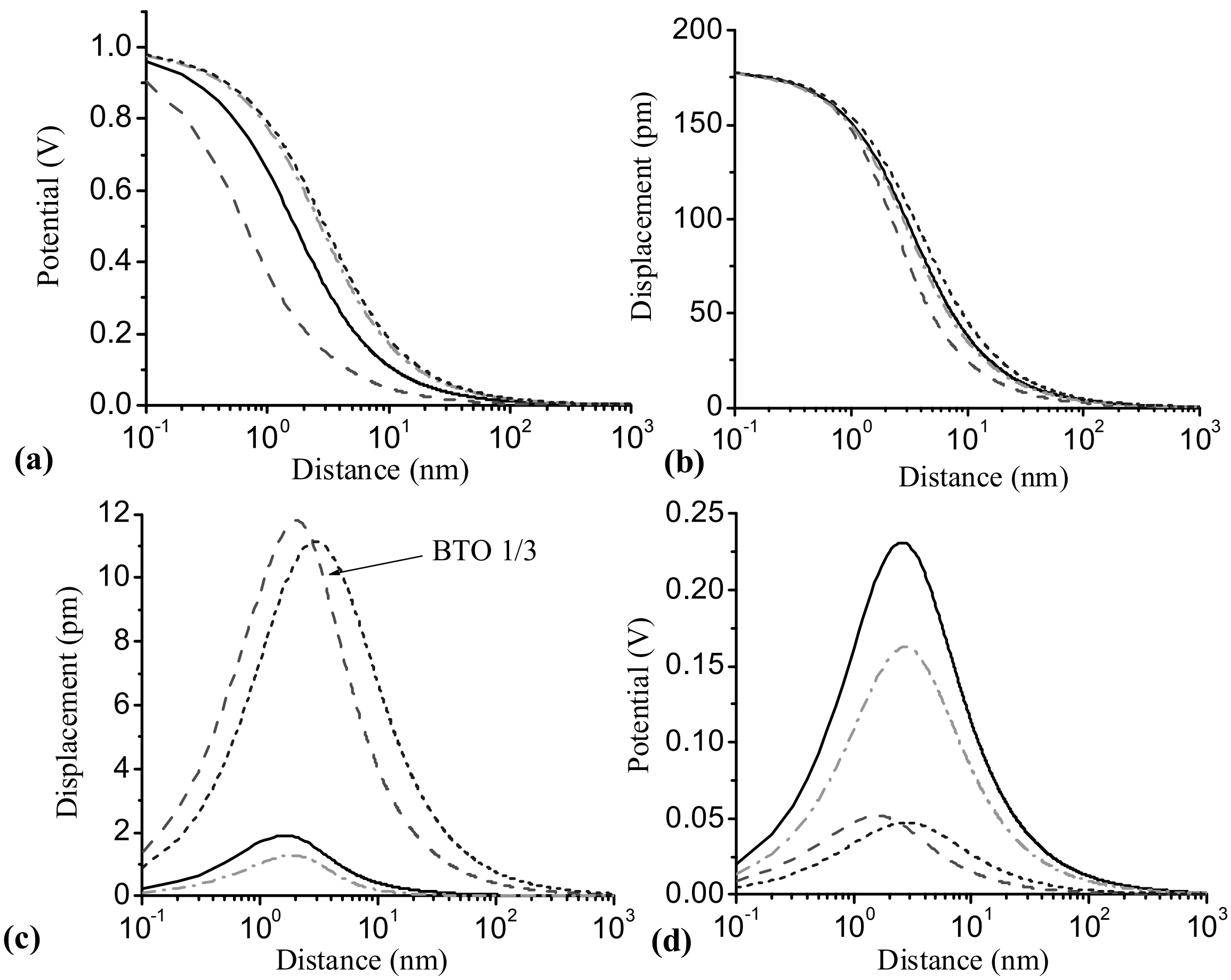

**Fig. 14.** S.V. Kalinin, E. Karapetian, and M. Kachanov

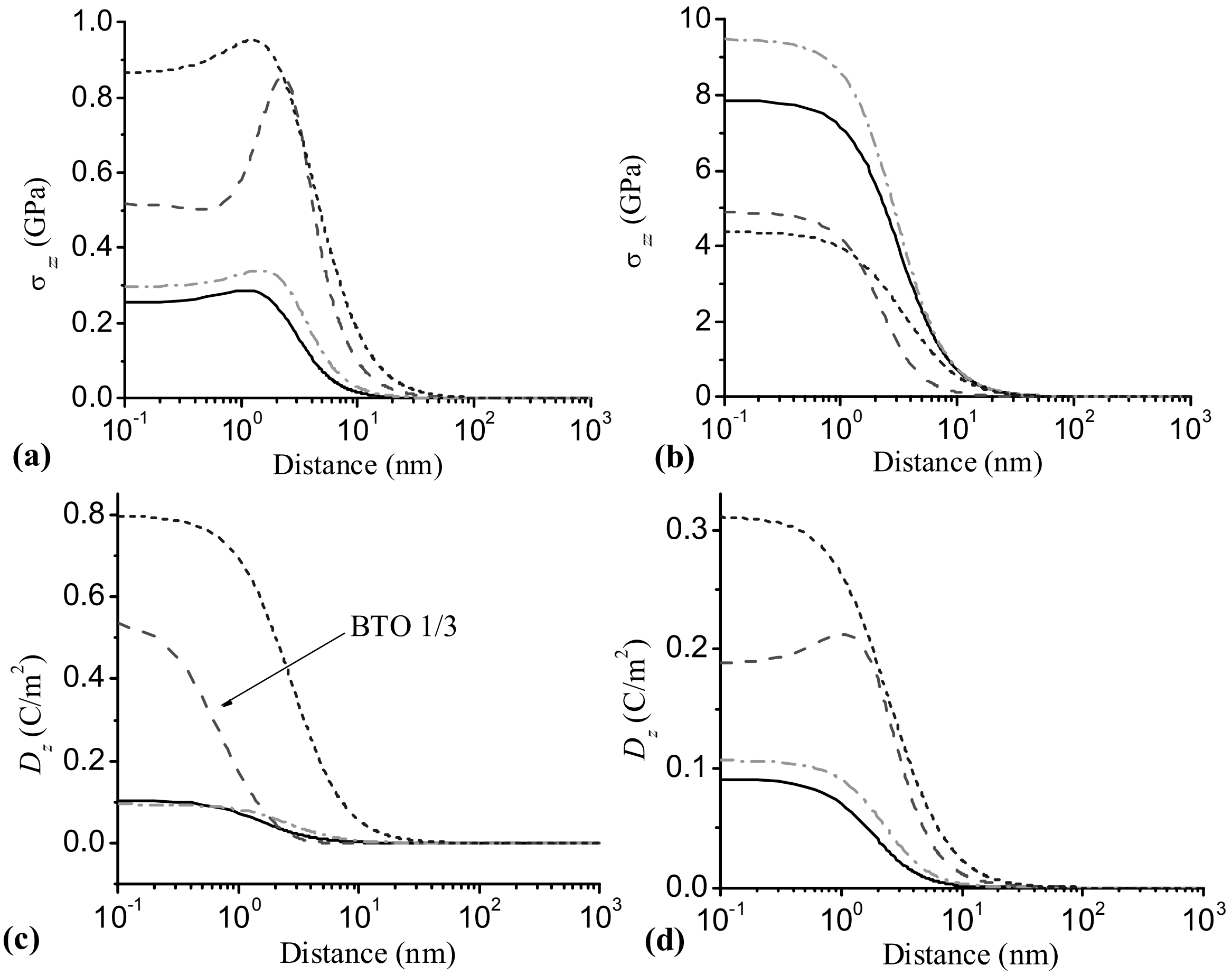

**Fig. 15.** S.V. Kalinin, E. Karapetian, and M. Kachanov

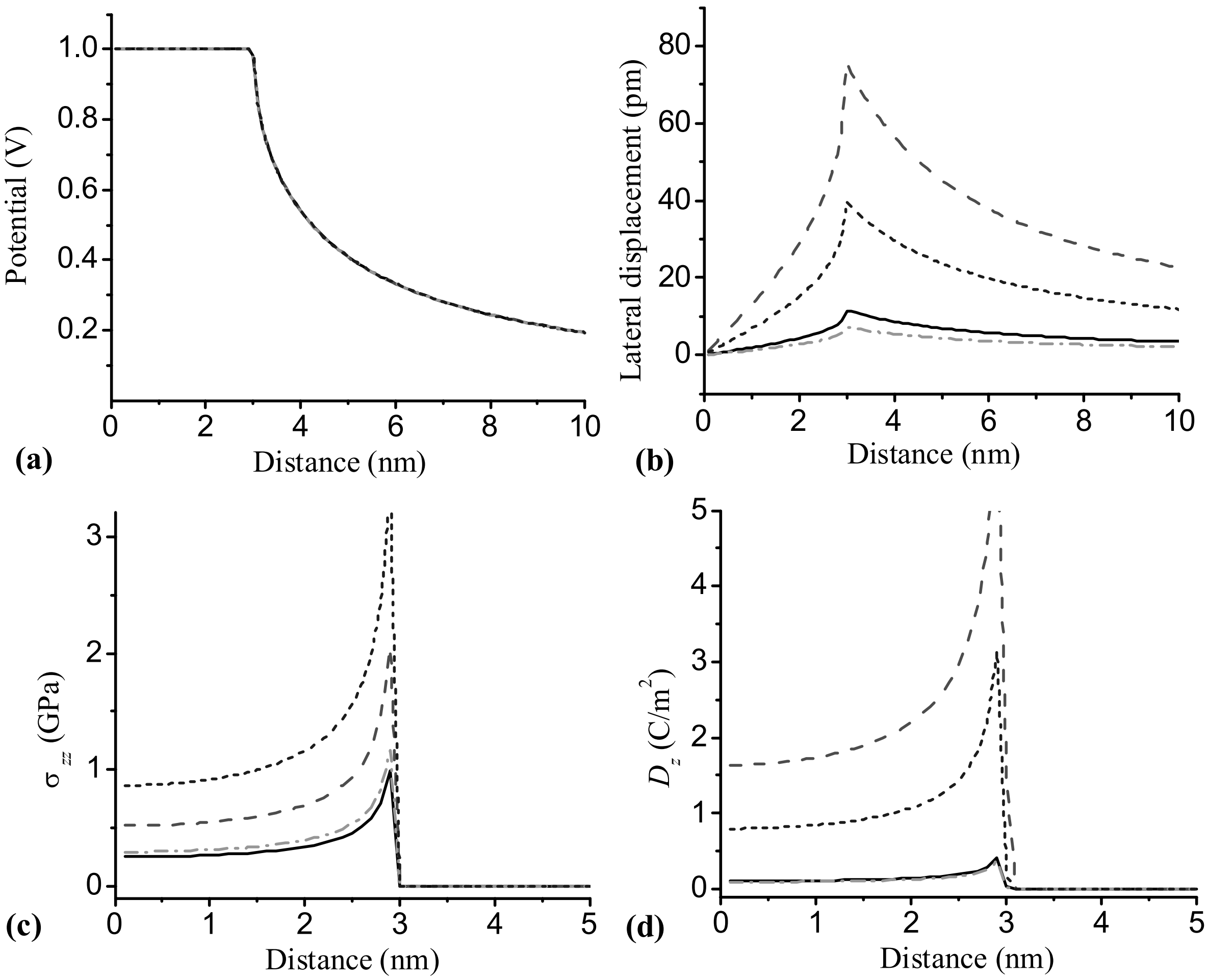

**Fig. 16.** S.V. Kalinin, E. Karapetian, and M. Kachanov

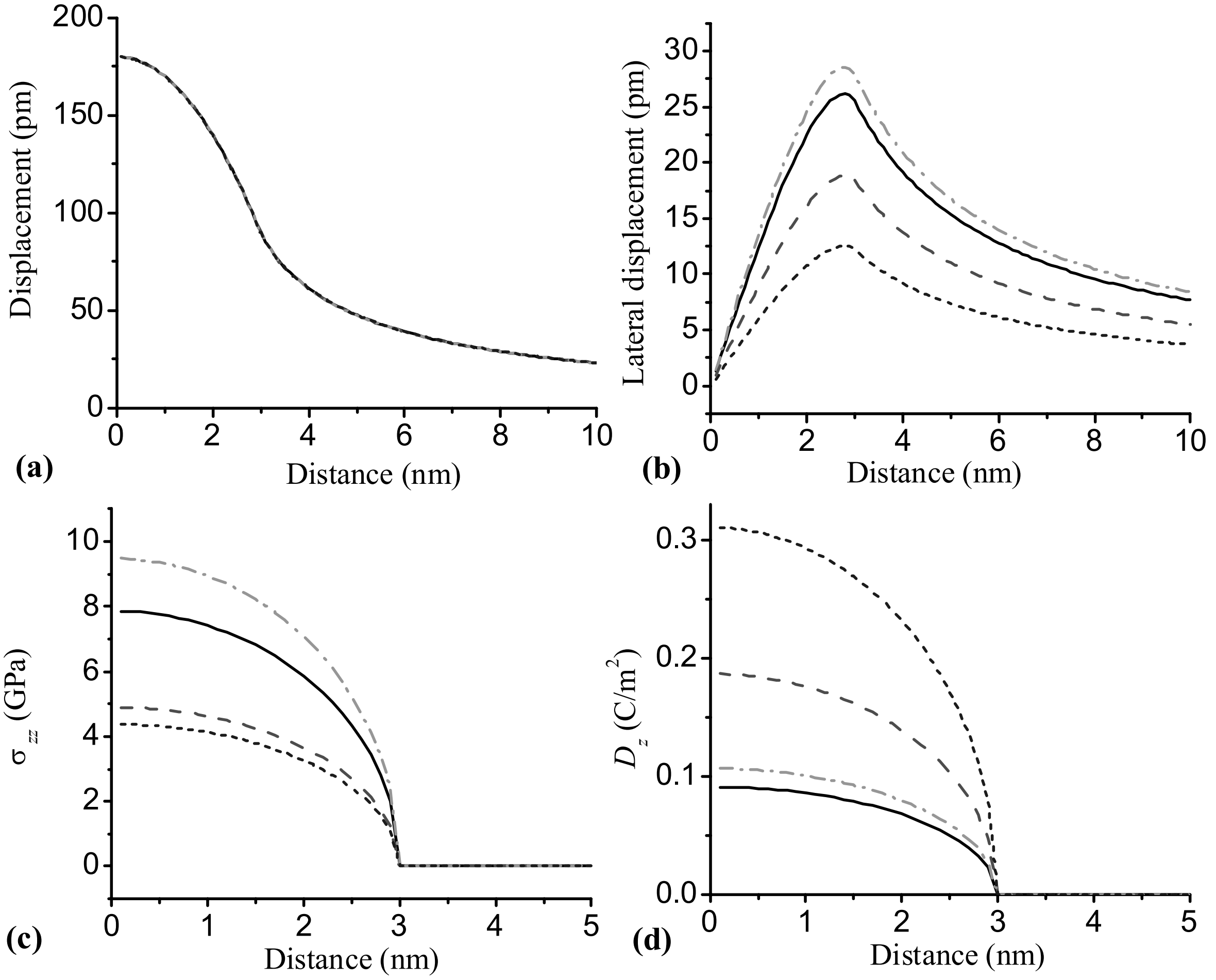

**Fig. 17.** S.V. Kalinin, E. Karapetian, and M. Kachanov

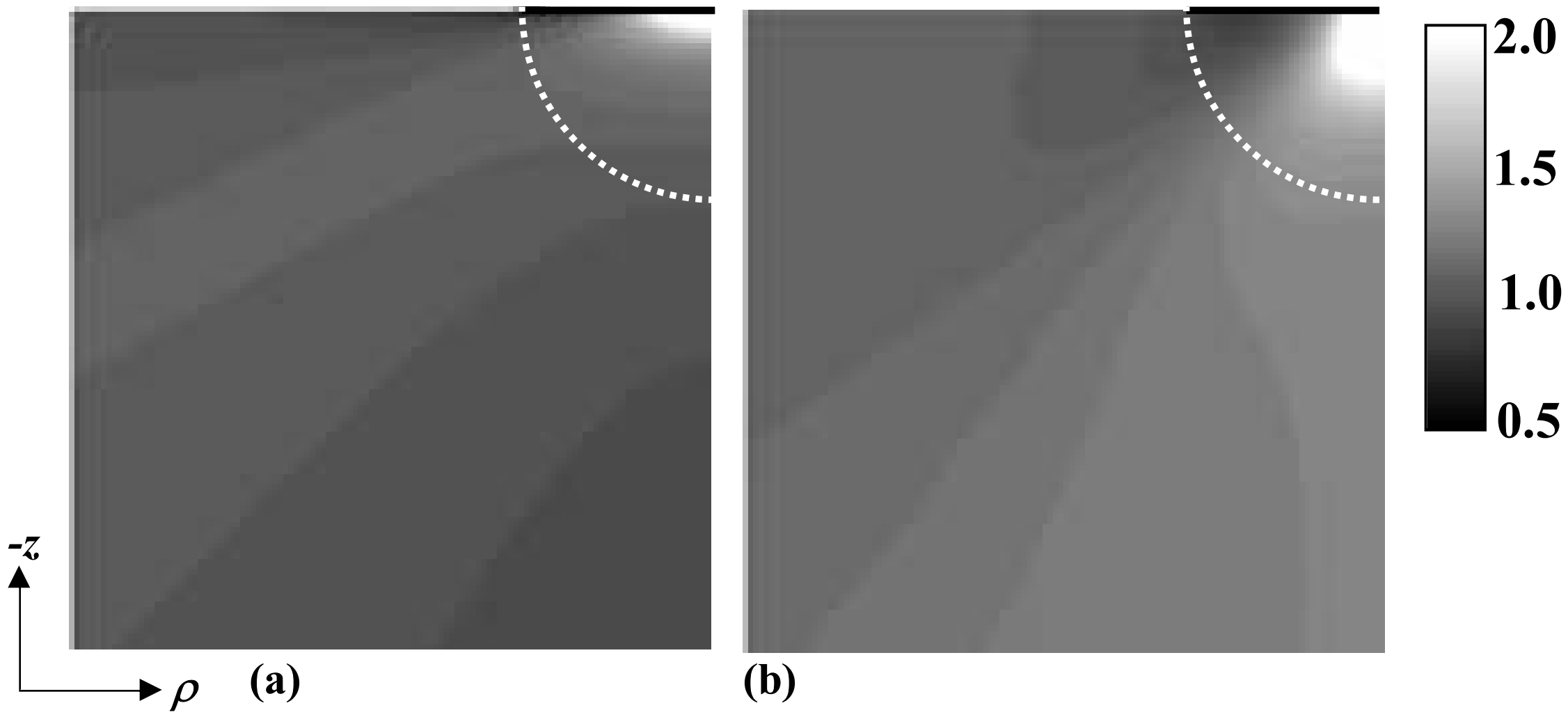

**Fig. 18.** S.V. Kalinin, E. Karapetian, and M. Kachanov

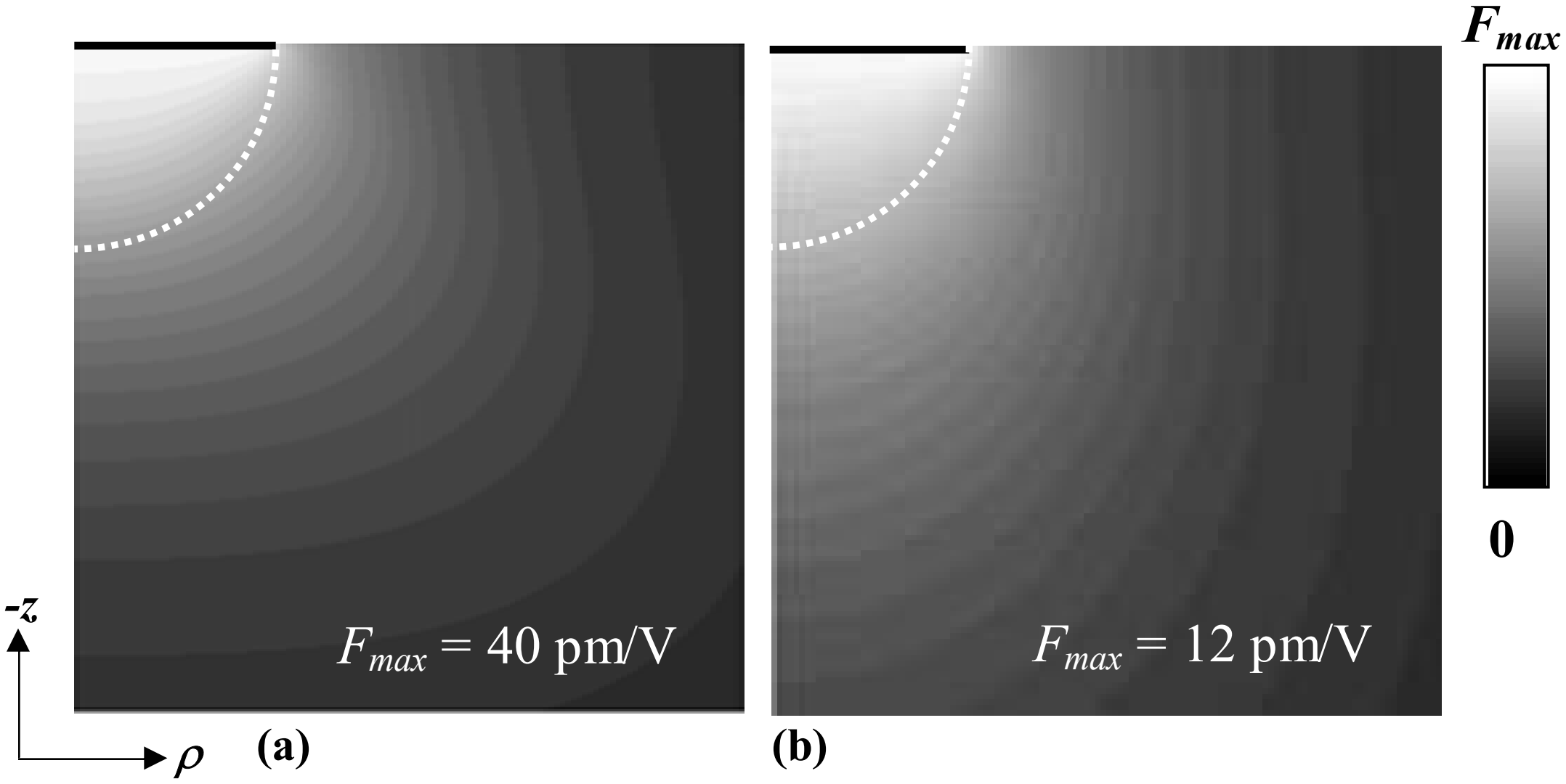

**Fig. 19.** S.V. Kalinin, E. Karapetian, and M. Kachanov